\documentclass[aps,reprint,nofootinbib,nobibnotes,notitlepage,superscriptaddress,twocolumn,prd,amsmath,amssymb
]{revtex4-2}

\usepackage[T1]{fontenc}
\usepackage{aecompl}
\usepackage{natbib}
\usepackage{verbatim}
\usepackage[usenames,dvipsnames]{xcolor}
\usepackage{tabulary}
\usepackage{tabularx}
\usepackage{todonotes}
\usepackage{hyperref}
\hypersetup{
	colorlinks = true,
    linkcolor = blue,
    urlcolor  = blue,
    citecolor = blue
}
\usepackage{braket}
\usepackage{float}
\usepackage{dcolumn}
\usepackage{bm}
\usepackage{microtype}
\usepackage{wasysym}
\usepackage{fontawesome5}
\newcommand{\aap}{A\&A}
\newcommand{\apjs}{ApJS}
\newcommand{\aj}{A.J.}
\newcommand{\mnras}{MNRAS}

\newcommand{\pasp}{PASP}

\newcommand{\Mh}{{M_\mathrm{h}}}

\newcommand{\abacussummit}{{\tt AbacusSummit}}

\numberwithin{equation}{section}

\defcitealias{Asgari2021}{A21}
\defcitealias{Hildebrandt2021}{H21}
\defcitealias{Takahashi2017}{T17}
\defcitealias{Heydenreich2023}{H23}
\defcitealias{Planck2020}{PL18}

\newcommand{\ttheta}[1]{\pmb{\theta_\mathrm{ap}}}

\newcommand{\Norm}[2]{\mathcal{N}}

\newcommand{\Gpch}{\,h^{-1}\text{Gpc}}

\newcommand{\omegac}{\omega_{\rm cdm}}
\newcommand{\omegab}{\omega_{\rm b}}

\newcommand{\Neff}{N_{\rm eff}}
\newcommand{\Msunh}{\,h^{-1}{\rm M_{\odot}}}

\newcommand{\nrun}{{\rm d}\, n_\mathrm{s}/{\rm d}\ln k}

\newcommand\be{\begin{equation}}
\newcommand\ee{\end{equation}}
\def\bea{\begin{eqnarray}}
\def\eea{\end{eqnarray}}

\allowdisplaybreaks
\begin{document}

\title{Constraining cosmological parameters using density split lensing\\ and the conditional stellar mass function}

\author{Pierre A. Burger}
\email[Corresponding author: ]{pierre.burger@uwaterloo.ca}
\affiliation{Waterloo Centre for Astrophysics, University of Waterloo, Waterloo, ON N2L 3G1, Canada}
\affiliation{Department of Physics and Astronomy, University of Waterloo, Waterloo, ON N2L 3G1, Canada}
\author{Darshak A. Patel}
\affiliation{Department of Physics and Astronomy, University of Waterloo, Waterloo, ON N2L 3G1, Canada}
\author{Michael J. Hudson}
\affiliation{Department of Physics and Astronomy, University of Waterloo, Waterloo, ON N2L 3G1, Canada}
\affiliation{Waterloo Centre for Astrophysics, University of Waterloo, Waterloo, ON N2L 3G1, Canada}
\affiliation{Perimeter Institute for Theoretical Physics, Waterloo, ON N2L 2Y5, Canada}


\date{\today}


\label{firstpage}
\begin{abstract}
In this work, we develop a simulation-based model to predict the excess surface mass density (ESD) depending on the local density environment. Using a conditional stellar mass function, our foreground galaxies are tailored toward the bright galaxy sample of the early data release of the Dark Energy Spectroscopic Instrument (DESI). Due to the nature of the ESD measurement, our derived model is directly applicable to all DESI data.
To build this model, we use the \texttt{AbacusSummit} N-body simulation suite from which we measure all necessary statistics and train an emulator based on \texttt{CosmoPower}. Finally, we present a cosmological parameter forecast for a possible combined analysis of DESI and the Ultraviolet Near Infrared Optical Northern Survey. 
\end{abstract}

\pacs{Valid PACS appear here}

\maketitle

\section{Introduction}
\label{sec:introduction}

To infer cosmological parameters of the standard model of cosmology, called the $\Lambda$ Cold Dark Matter model ($\Lambda$CDM), previous works made use of the density-split statistics (DSS), which extracts information based on environments measured from smoothing foreground galaxy samples \cite{Gruen:2018,Burger2023,Paillas2023,Burger2024b}. Afterwards, these environments are assigned to different quantiles, which in turn are used to correlate with either foreground galaxies' positions or background galaxies' shear. This work focuses on the correlation with shear, which we call density split lensing (DSL). In \cite{Gruen:2018,Burger2023}, cosmological and galaxy-halo parameters were inferred using the DSL, and they have shown that the DSL is a competitive method compared to the usually used second-order statistic. The reason is that second-order statistics can only extract the Gaussian information from the data. In contrast, DSS is a higher-order statistic and can extract information beyond the Gaussian part of the data. The potential power of higher-order statistics was nicely demonstrated in \cite{HOWLS2023} for weak lensing statistics and in \cite{Beyond2ptCollaboration2024} for galaxy clustering statistics.

Although analytical models of the DSL \cite{Friedrich:2018, Burger2022} provide insights into the underlying properties of the large-scale structure (LSS) they rely on several assumptions, like the galaxy bias, that are only valid if large scales are considered. In contrast, simulation-based models have the drawback that they are measured from expensive simulations and, therefore, depend on the noise of the simulations and the number of training nodes used to train the emulator. However, simulation-based models enable a more complicated connection between dark matter haloes and galaxies. A possible approach to model the galaxy-halo connection is using a Halo Occupation Distribution (HOD) \cite{Peacock2000}, which populates the dark matter haloes with central and satellite galaxies depending on the halo's mass. Using a HOD instead of linear galaxy bias, which is only valid on large scales, allows using smaller scales. 

Many previous studies have modelled the spatial distribution of galaxies of a specific class (e.g.\ Luminous Red Galaxies in Baryon Oscillation Spectroscopic Survey \citep[BOSS;][]{Dawson2013}), for which a simple HOD model that predicts the \emph{number} of galaxies per halo is sufficient. Future galaxy data sets, such as the DESI bight galaxy sample (BGS) \citep{Ruiz2020,Hahn2023A}, will instead contain galaxies with a range of luminosity and stellar mass. While one could bin these galaxies by stellar mass and fit an HOD to each bin separately, it's better to model all stellar masses simultaneously. An extension of the HOD that does this is the Conditional Stellar Mass Function (CSMF) model \cite{Cooray2005,Yang2008,Cacciato2009,Cacciato2013,Dvornik2023}.

Similar to \cite{Burger2024b}, we follow the idea of \cite{Cuesta-Lazaro2023,Paillas2023,Burger2024b} and build a model based on the \texttt{AbacusSummit}\footnote{\url{https://abacussummit.readthedocs.io}} simulations \citep{Maksimova2021}. Using the halo catalogues of the \texttt{AbacusSummit} simulations, we built a CSMF HOD that predicts the number of central and satellites per halo and the stellar mass of those galaxies\footnote{A current implementation of the CSMF HOD will be made public here \url{https://github.com/abacusorg/abacusutils}.}. Using those galaxies, we define environments that depend on the galaxy number density and measure those environments' excess surface mass density (ESD). Based on different cosmologies and CSMF parameters, we built an emulator for the ESD, which can predict the model everywhere in the predefined parameters space. Emulators are a machine learning tool widely used in cosmology due to their speed and accuracy in interpolating the model vector between predefined training nodes \cite{2020MNRAS.491.2655M,COSMOPOWER2022,EE2021,Angulo2021,Bonici2024}. Furthermore, while we used the projected environment in \cite{Burger2024b}, we follow here the work of \cite{Paillas2023} and define the environment in the three-dimensional space, which is only possible because we tailor our foreground galaxies towards the spectroscopic data from early data release of the Dark Energy Spectroscopic Instrument (DESI-EDR). In future work, we expect to measure the ESD from the Ultraviolet Near-Infrared Optical Northern Survey (UNIONS) \citep{Guinot2022}, so here we take this as a reference survey for the background sources.

A natural extension to the work presented here would be to perform a similar analysis as in \cite{Burger2024} and combine DSL with density split clustering as performed in \cite{Paillas2023}. As shown in \cite{Burger2024}, combining clustering and lensing statistics improves constraining cosmological and HOD parameter constraints. We ignored it for this work because we have seen in \cite{Burger2024} that clustering statistics are significantly more challenging to emulate due to the small uncertainties. We therefore use only the galaxy-galaxy lensing signal for lens galaxies in different environments combined with predictions of the stellar mass function (SMF).

This work is structured as follows. In Sect.~\ref{Sec:ESD}, we review the basics of the excess surface mass density (ESD). In Sect.~\ref{sec:data}, we briefly describe the reference data used to define the prior range and create a realistic covariance matrix for a parameter forecast and validate the emulator's accuracy. In Sect.~\ref{sec:AbacusSummit}, we review the basics of the \texttt{AbacusSummit} simulations that we use to build our model and the implementation of the CSMF. In Sect.~\ref{sec:T17}, we describe our numerical covariance matrix estimation using the simulations described in \cite[][thereafter T17]{Takahashi2017}. In Sect.~\ref{sec:emulator}, we describe the accuracy of our emulator model, and in Sect.~\ref{sec:forecast}, we perform a forecast and finalize our work in Sect.~\ref{sec:conclusion}.

\section{Excess surface mass density}
\label{Sec:ESD}

This work focuses on measuring the ESD of a geometrical thin lens, which is defined assuming circular symmetry as
\begin{equation}
    \Delta \Sigma (r_\mathrm{p}) = 
    \frac{2}{r_\mathrm{p}^2} \int_0^{r_\mathrm{p}} \Sigma (R) \ R \ \mathrm{d}R- \Sigma (r_\mathrm{p}) \, 
\end{equation}
where $r_\mathrm{p}$ is the projected distance between the lens and the impact of the light ray in the lens plane. The surface mass density, $\Sigma (r_\mathrm{p})$, is determined by the line-of-sight integration along the comoving distance, $\chi$, of the mass density, $\rho(\mathbf{r_\mathrm{p}},\chi)$, inside the geometrical thin lens with lower bound, $\chi_\mathrm{low}$, and upper bound, $\chi_\mathrm{up}$,
\begin{equation}
    \Sigma (r_\mathrm{p}) = 2\pi \int_{\chi_\mathrm{low}}^{\chi_\mathrm{up}} \rho(|\mathbf{r_\mathrm{p}}|,\chi) \ \mathrm{d}\chi \ .
\end{equation}
The ESD is closely related to the tangential shear, $\gamma_\mathrm{t}$, through
\begin{equation}
    \Delta \Sigma(r_\mathrm{p}) = \gamma_\mathrm{t}(r_\mathrm{p}) \ \Sigma_\mathrm{crit}(\chi_\mathrm{l},\chi_\mathrm{s}) \ ,
\end{equation}
where $\Sigma_\mathrm{crit}$ is a geometrical factor called the critical surface mass density and is defined as
\begin{equation}
    \Sigma_\mathrm{crit}(\chi_\mathrm{l},\chi_\mathrm{s}) = \frac{c^2}{4 \pi G} \frac{D_\mathrm{A}(\chi_\mathrm{s})}{D_\mathrm{A}(\chi_\mathrm{l}) \ D_\mathrm{A}(\chi_\mathrm{s}-\chi_\mathrm{l})} \ ,
\end{equation}
where $D_\mathrm{A}$ denotes the angular diameter distance, $\chi_\mathrm{l}$ the comoving distance to the lens plane and $\chi_\mathrm{s}$ the comoving distance to the source which emitted the light.

A simple minimum variance estimator of the ESD is given by averaging over many sources weighted by critical surface density
\begin{equation}
    \Delta \Sigma (r_\mathrm{p}) = \frac{\sum_{ls} \Sigma_\mathrm{crit}^{-1}(\chi_\mathrm{l},\chi_\mathrm{s})\ \gamma_\mathrm{t}(r_\mathrm{p})}{\sum_{ls} \Sigma_\mathrm{crit}^{-2}(\chi_\mathrm{l},\chi_\mathrm{s})} \ , 
\end{equation}
where $\sum_{ls}$ is the sum over all lens source pairs of a bin with center $r_\mathrm{p}$. 
Since the ESD dependency on the source distance distribution is cancelled out by the $\Sigma_\mathrm{crit}$ division, we used a single redshift slice to measure the ESD in the simulations. The measured ESDs with that approach give unbiased results when compared to real data ESD measurements, as long as the redshift distribution of the real data is behind the redshift distribution of the lens \cite{Lange2024}. Therefore, this allows us to apply the derived model in this work to real data measurements without any modifications in the future. Although the so-called boost correction could potentially correct for an overlapping lens and source redshift distribution, a well-separated lens and source redshift distribution also avoid the complicated modelling of intrinsic alignments and magnification effects.

\section{Reference data}
\label{sec:data}

\begin{figure}
\includegraphics[width=\linewidth]{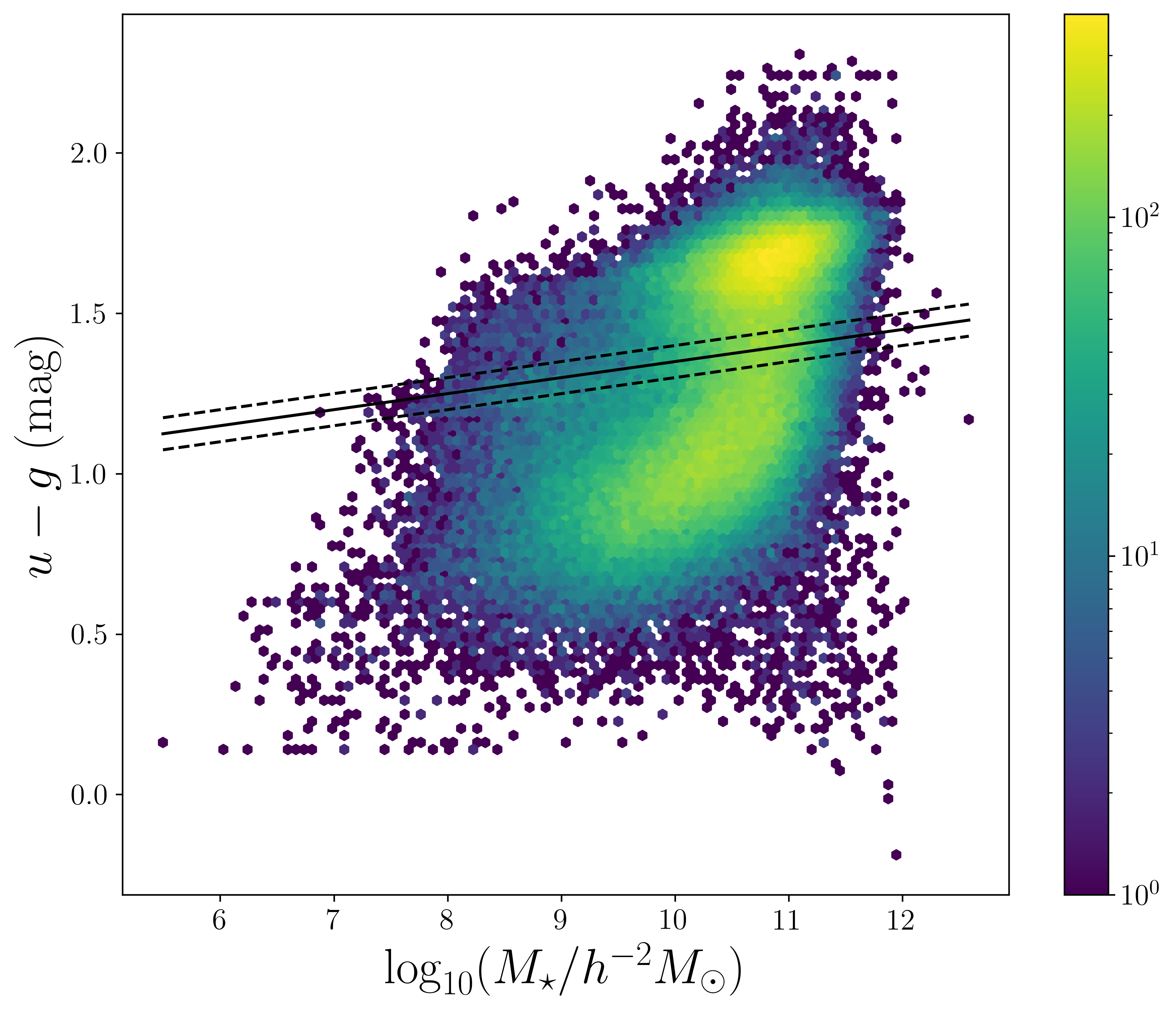}
\caption{Color versus stellar mass diagram. The solid line cuts right through the green valley, separating red and blue galaxies. To avoid interlopes, we shifted the cut criteria slightly upwards.}
\label{fig:color}
\end{figure}

For this work, we develop a model tailored towards a galaxy-galaxy analysis using the galaxy positions from (DESI) \cite[][]{DESI2024A,Adame2024} as lenses and the galaxy positions and shapes from the UNIONS as sources. 

UNIONS is a weak lensing survey that measures galaxy shapes and photometric redshifts above a declination of $+30^\circ$ and a galactic latitude $|b|>25^\circ$. In this work, the information we need to derive the model is the mean redshift $z\approx 0.65$ of the galaxies, which will be used to calibrate photometric redshift distributions. Besides this, we also use the shape measurements from the preliminary UNIONS catalogue (v1.3 ShapePipe) to estimate a covariance matrix, which we further discuss in Sect.~\ref{sec:T17}.

DESI is a spectroscopic survey that seeks to map the LSS of the Universe by measuring the positions and redshifts of 30 million pre-selected galaxies across one-third of the night sky \cite{DESIoverview2022}. In this work, we make use of the BGS sample \cite[][]{Ruiz2020,Hahn2023A} of the \texttt{sv3} early-data release of DESI \cite{DESIEDR2024}, where \texttt{sv3} is a $140\,\mathrm{deg}^2$ of DESI using the same targeting as the main survey with a high fibre assignment completeness to enable unbiased small scale clustering measurements. In particular, we are using the clustering catalogues \texttt{BGS\_BRIGHT\_N\_clustering} and \texttt{BGS\_BRIGHT\_S\_clustering}, described in \cite{Ross2024}. In addition, we extract the stellar masses and absolute magnitudes from the Legacy Survey DR9 in the $u,g,r$-bands \cite{Zou2017,Dey2019} from the \texttt{fastspec-fuji-sv3-bright} catalogue. These stellar masses are determined using the \texttt{fastspecfit} software described in \cite{Moustakas2023}. Using the stellar masses and the $u-g$ colour of the galaxies, we select only red galaxies by applying the following criteria 
\begin{equation}
    u-g > 0.05 \log_{10}(M_\star/M_\odot) + 0.9 \ ,
\end{equation}
where the parameters are chosen such that we cut through the green valley of the galaxy sample, as seen in Fig.~\ref{fig:color}. Using this selection, we decreased the number of galaxies from 143853 to 62244. This sample of galaxies is used to estimate the SMF, which in turn is used to define the prior range of our HOD parameters.

\begin{figure}
\includegraphics[width=\linewidth]{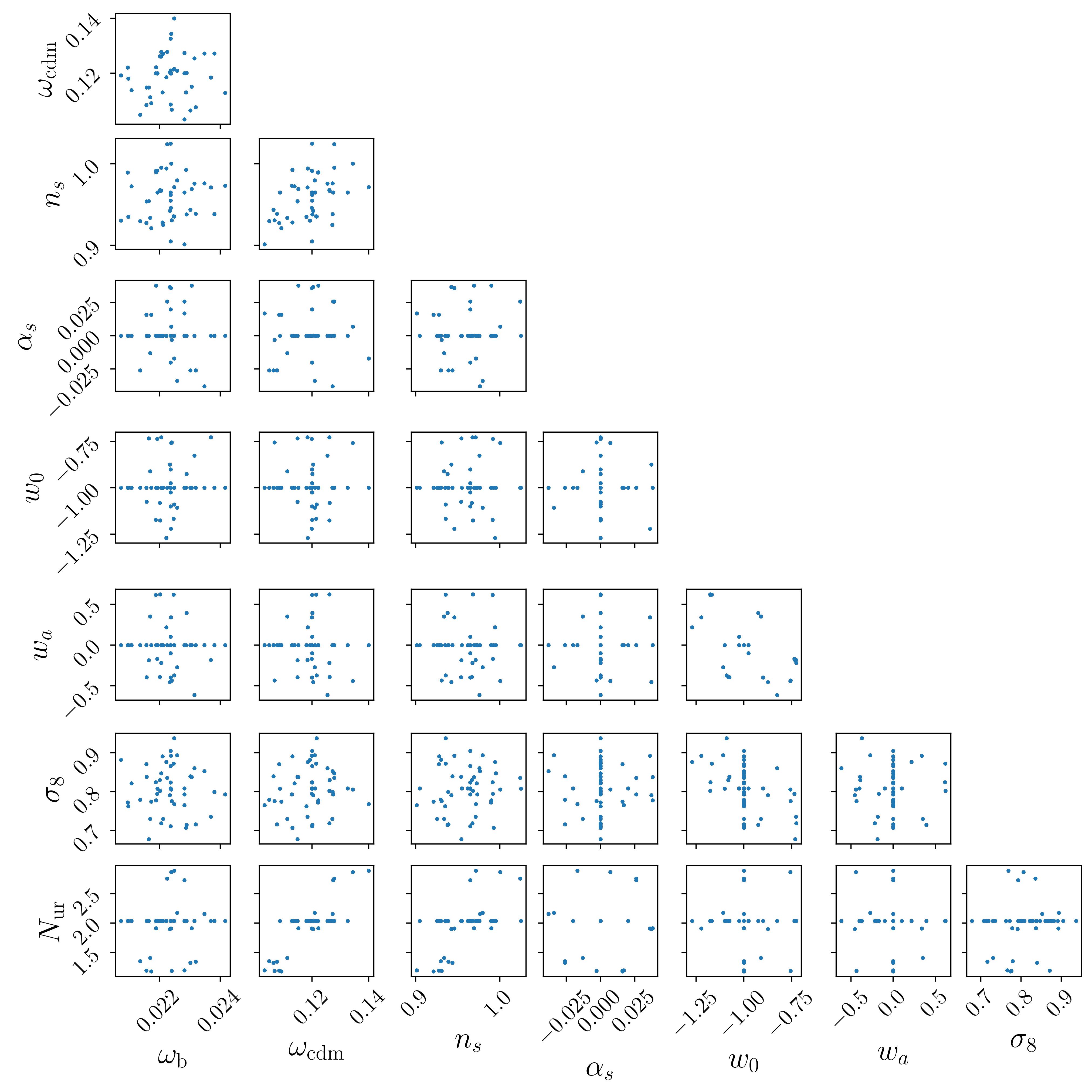}
\caption{Parameter distribution of all available \texttt{AbacusSummit} cosmologies.}
\label{fig:abacus_paramters}
\end{figure}

\section{AbacusSummit light cone simulations}
\label{sec:AbacusSummit}

This work uses the \abacussummit\ \citep{Maksimova2021} to build an emulator model to predict the ESD. Here, we mention only the essential details to follow the work of this paper. The \abacussummit\ is a cosmological simulation run with the \texttt{Abacus} N-body code \citep{Garrison2019, Garrison2021}. The \textit{base} \abacussummit\ simulations follow the evolution of $6912^3$ dark matter particles in a periodic box with a side length of $2\Gpch$, and a mass resolution of $2.1 \times 10^9 \Msunh$. Group or haloes of particles are found with the \texttt{CompaSO} hybrid Friends-of-Friends/Spherical Overdensity algorithm \citep{Hadzhiyska2022}. Lightcone simulations are described in \cite{Hadzhiyska2022}, which we use to construct realistic mock catalogues of the sky that vary with redshift. The light cones cover an octant of the sky up to a redshift of 0.8\footnote{At large redshift the octant decreases due to the placement of the boxes to build the light cones.} and span 63 different cosmologies within an extended $w_0w_a$CDM parameter space (see Fig.~\ref{fig:abacus_paramters}),
\begin{equation}
\label{eq:abacus_parameters}
    \Theta_{\rm \texttt{AbacusSummit}} = \{ \omegac, \omegab, \sigma_8, n_\mathrm{s}, \nrun, \Neff, w_0, w_a \} \,,
\end{equation} 
where $\omegac$ and $\omegab$ are the physical cold dark matter and baryon densities, $\sigma_8$ is the amplitude of density fluctuations, $n_\mathrm{s}$ is the spectral index of the primordial power spectrum, $\nrun$ is the running of the spectral index, $\Neff$ is the effective number of ultra-relativistic species, $w_0$ is the present-day dark energy equation of state, and $w_a$ captures the evolution of dark energy across redshift. The simulations assume zero curvature, and the value of the Hubble parameter $h$ is set by matching the Cosmic Microwave Background angular size of the acoustic scale, $\theta_*$, as observed by \citetalias{Planck2020}. For the fiducial cosmology \texttt{c000}, 25 phase realizations with different initial conditions are available, which we later use to correct for cosmic variance \cite{Racz2023}. Only one phase realization, \texttt{ph00}, is available for all the other cosmologies.

We use the reduced shear information, $g=\gamma/(1-\kappa)$, a single redshift slice $z=0.65$ to measure the ESD described above. The chosen redshift is roughly the median redshift of the calibration sample used to calibrate the photometric redshift of the UNIONS data. The resolution of this shear map is $A_\mathrm{pix}=0.046\,\mathrm{arcmin}^2$ $(\texttt{nside}=16384)$, or a side length of $49.4\,h^{-1}\mathrm{kpc}$ at a redshift of $z=0.4$ given by the \citetalias{Takahashi2017} cosmology.

\subsection*{Galaxy-halo connection model}
\label{subsec:hod}

To model the connection between dark matter haloes and galaxies we use a CSMF \cite{Cooray2005,Yang2008,Cacciato2009,Cacciato2013,Dvornik2023}. The CSMF can be divided into the contributions from central (c) and satellite (s) galaxies,
\begin{equation}
    \Phi(M_\star|\Mh) = \Phi_\mathrm{c}(M_\star|\Mh) + \Phi_\mathrm{s}(M_\star|\Mh) \ .
\end{equation}
We model the CSMF of the central galaxies using a log-normal distribution
\begin{equation}
    \Phi_\mathrm{c}(M_\star|\Mh) = \frac{1}{\sqrt{2\pi}\ln(10)\sigma_\mathrm{c} M_\star} \exp\left[ -\frac{\log_{10}(M_\star/M_\mathrm{c}^*)^2}{2\sigma_\mathrm{c}^2}\right] \ ,
\end{equation}
and the CSMF from the satellite galaxies with a modified Press-Schechter function
\begin{equation}
    \Phi_\mathrm{s}(M_\star|\Mh)  = \frac{\phi_\mathrm{s}}{M_\mathrm{s}^*} \left(\frac{M_\star}{M_\mathrm{s}^*}\right)^\alpha \exp\left[-\left(\frac{M_\star}{M_\mathrm{s}^*}\right)^2\right] \, 
\end{equation}
where $\sigma_\mathrm{c}$ is a free parameter describing the scatter between the stellar and halo mass. The $\alpha$ parameter in general, depends on the halo mass $\Mh$ and governs the power law behaviour of satellite galaxies as given by
\begin{equation}
    \alpha(\Mh) = -2.0 + a_1(1-\arctan \left[a_2\log_{10}(M_\star/M_2)\right]) \ .
\end{equation}
However, for this work, we followed \cite{Dvornik2023} and set $a_2=0$, eliminating the $M_2$ parameter, too.
This model statistically assigns galaxies to haloes based on their properties, most importantly, the host halo mass.
We parametrize the remaining parameters as:
\begin{equation}
    M_\mathrm{c}^*(\Mh) = M_0 \frac{(\Mh/M_1)^{\gamma_1}}{(1+\Mh/M_1)^{\gamma_1-\gamma_2}} \ , 
\end{equation}
\begin{equation}
    M_\mathrm{s}^*(\Mh) = 0.56 M_\mathrm{c}^*(\Mh) \ ,
\end{equation}
and 
\begin{equation}
    \log_{10}[\phi_\mathrm{s}(\Mh)] =  b_0 + b_1 \log_{10}(M_{12}) + b_2 [\log_{10}(M_{12})]^2 \ ,
\end{equation}
where we have set $b_2=0$ for this work, and $M_{12} = \Mh/(10^{12}\, h^{-1} M_\odot)$. This leaves us with eight free parameters to model the galaxy-halo connection $\{M_0, M_1, \gamma_1, \gamma_2, \sigma_\mathrm{c}, a_1, b_0, b_1\}$.

Given the CSMF, the expectation values of centrals and satellites in the range $M_{\star,1} < M_\star < M_{\star,2}$ is given by
\begin{align}
    \langle N_{\rm X} \rangle(\Mh) = \int_{M_\star,1}^{M_\star,2} \Phi_\mathrm{X}(M_\star|\Mh) \ \mathrm{d} M_\star \ .
\end{align}
Furthermore, the CSMF can also be used to predict the stellar mass function (SMF) 
\begin{equation}
\Phi_\mathrm{X}(M_\star) = \int_{0}^{\infty} \Phi_\mathrm{X}(M_\star|\Mh) \ n(\Mh) \ \mathrm{d} \Mh \ .
\label{eq:SMF}
\end{equation}

Motivated by \cite{Yuan2022, Burger2024b}, we extend the base model to account for environment-based secondary bias (assembly bias), captured by the parameters $B_{\rm cen}$ and $B_{\rm sat}$, which modifies the expectation of centrals and satellites depending on the local environment of the halo. As this is not done in the literature for a CSMF description, we found that modifying $M_1$ and $a_1$ as 

\begin{figure}
\includegraphics[width=\linewidth]{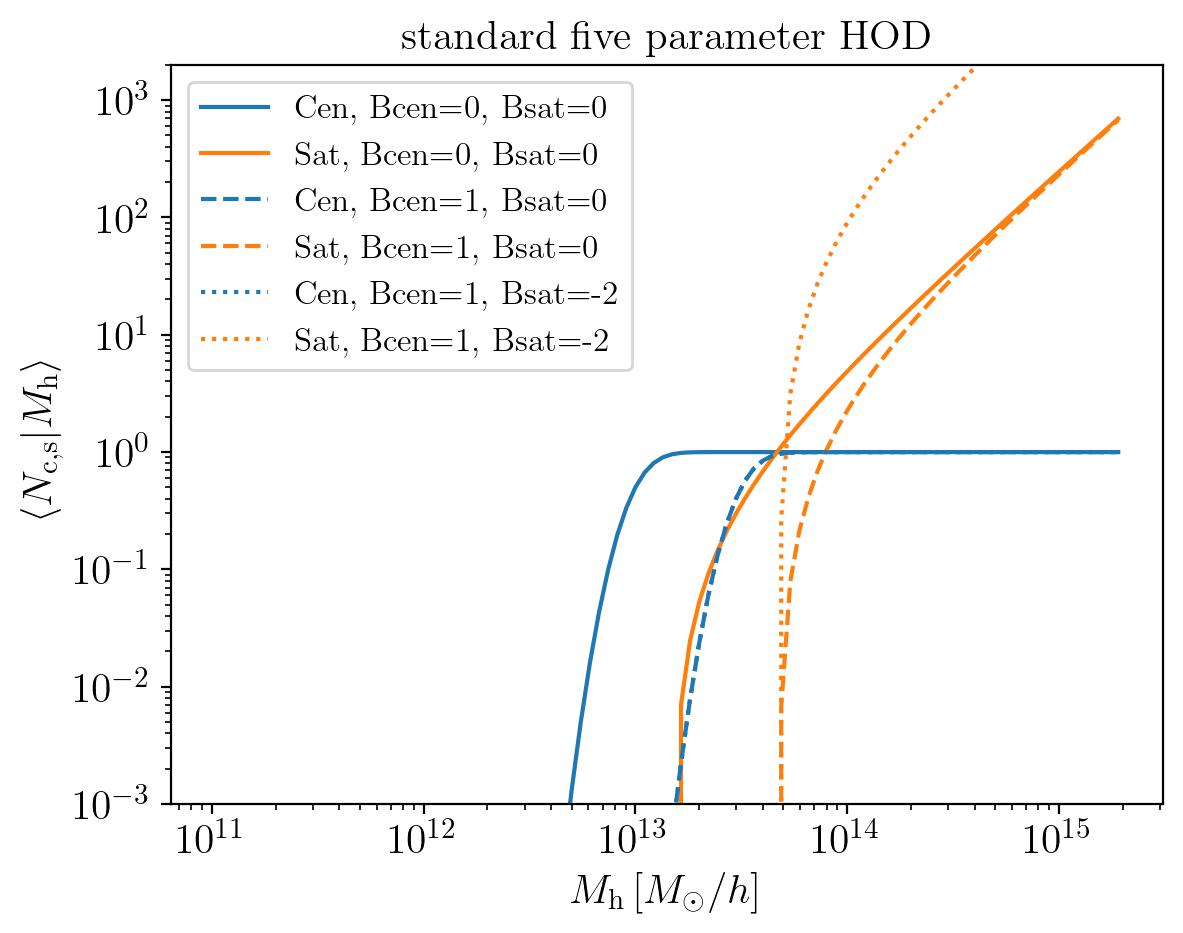}
\includegraphics[width=\linewidth]{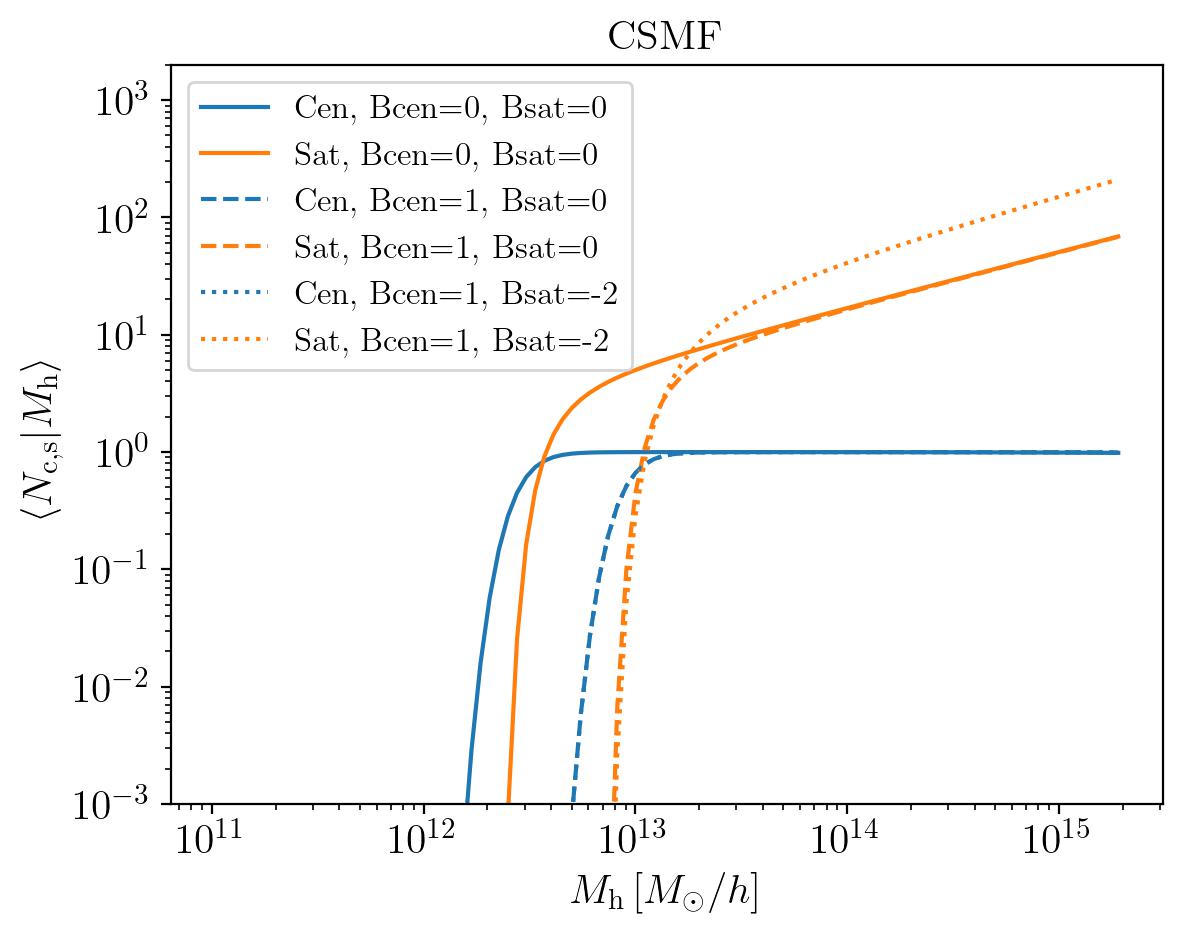}
\caption{Dependence of the expectation value on the environment-based secondary bias parameters. In the upper panel, we show the dependence if we use the standard HOD description used, for instance, for luminous red galaxies \cite{Burger2024b}. The lower panel shows that the modifications of the CSMF HOD described in this work result in similar effects. }
\label{fig:assmebly_bias}
\end{figure}

\begin{align}
    \mathrm{log}_{10} M_1^\mathrm{eff} &= \mathrm{log}_{10} M_1 + B_\mathrm{cen}(\delta^\mathrm{rank} - 0.5)\, , \\
    a_\mathrm{1}^\mathrm{eff} &= a_\mathrm{1} + B_\mathrm{sat}(\delta^\mathrm{rank} - 0.5) \, ,
\end{align}
results in outcomes similar to those used in \cite{Burger2024b} see Fig.~\ref{fig:assmebly_bias}. 
The local environment is defined by the smoothed matter density\footnote{Here we used a top-hat filter of radius $R_s = 3\,h^{-1}{\rm Mpc}$.} around the halo centres, while the halo itself is taken out of the measurement. Next these local environments are assigned to `ranks' $\delta^\mathrm{rank} \in [0,1]$, with environments denser than the median having $\delta^\mathrm{rank}>0.5$. 

As in \cite{Burger2024b}, we use the particles themselves to place satellites into the haloes. Originally, every particle had the same probability of hosting a satellite galaxy. We can relax this based on the distance to the center of the halo \citep{Yuan2018},
\begin{equation}
    p_i = \frac{\langle N_{\rm s} \rangle(M) }{N_\mathrm{p}}\left[1+s\left(1-\frac{2\,R_i}{N_\mathrm{p}-1}\right)\right]\, ,
\end{equation}
where $s$ is a free parameter to modulate this behaviour, $N_\mathrm{p}$ is the number of particles in the halo, $R_i$ is the rank of the particle ordered by distance to the halo centre\footnote{The outermost particle has rank 0, and the innermost particle has ranking $N_\mathrm{r}-1$.}.
As found in \cite{Debackere2021}, this modulation of the satellite profile can account for baryonic feedback processes. This leaves us finally with eleven free parameters to model the galaxy-halo connection $\{M_0, M_1, \gamma_1, \gamma_2, \sigma_\mathrm{c}, a_1, b_0, b_1, B_\mathrm{cen}, B_\mathrm{sat}, s\}$.

\section{Prior of the HOD parameters}
\label{sec:prior}
We noticed that most parameter combinations resulted in unrealistic scenarios if all eleven HOD parameters were freely varying. This includes cases where the number densities of galaxies and the satellite fraction were ridiculously low or high.

To constrain our prior range to only realistic parameter combinations, we decide on the following three criteria that every parameter combination must satisfy:
\begin{enumerate}
    \item The $\chi^2$ between SMF of the red galaxies of DESI-EDR and the SMF measured with halo mass function from \abacussummit\ \texttt{c000} must be smaller than 100. We elaborate more on this further down.
    \item The satellite fraction is at least 10$\%$ for stellar masses $10.5 < \log_{10} [M_\star/(h^{-1}\mathrm{M}_\odot)] < 10.8$, 5$\%$ for stellar masses $10.8 < \log_{10} [M_\star/(h^{-1}\mathrm{M}_\odot)] < 11.2$ and 1$\%$ for stellar masses $11.2 < \log_{10} [M_\star/(h^{-1}\mathrm{M}_\odot)] < 12$. These bins are used in our companion paper (Patel in prep.), and the lower limits adopted here are typically a factor of 4 lower than the best-fit values found in \cite{Hudson2015}. 
    \item At most $0.3\%$ of the available particles are used to distribute satellites.
\end{enumerate}

The context regarding criterion 3 is that we save, due to storage issues, only $0.9$\% of the particles available. To ensure that the satellite profile and the assembly bias modulation are possible, we ensured that none of the parameters requires more than $\sim 0.3\%$ of the available particles. 

\begin{figure}
\includegraphics[width=\linewidth]{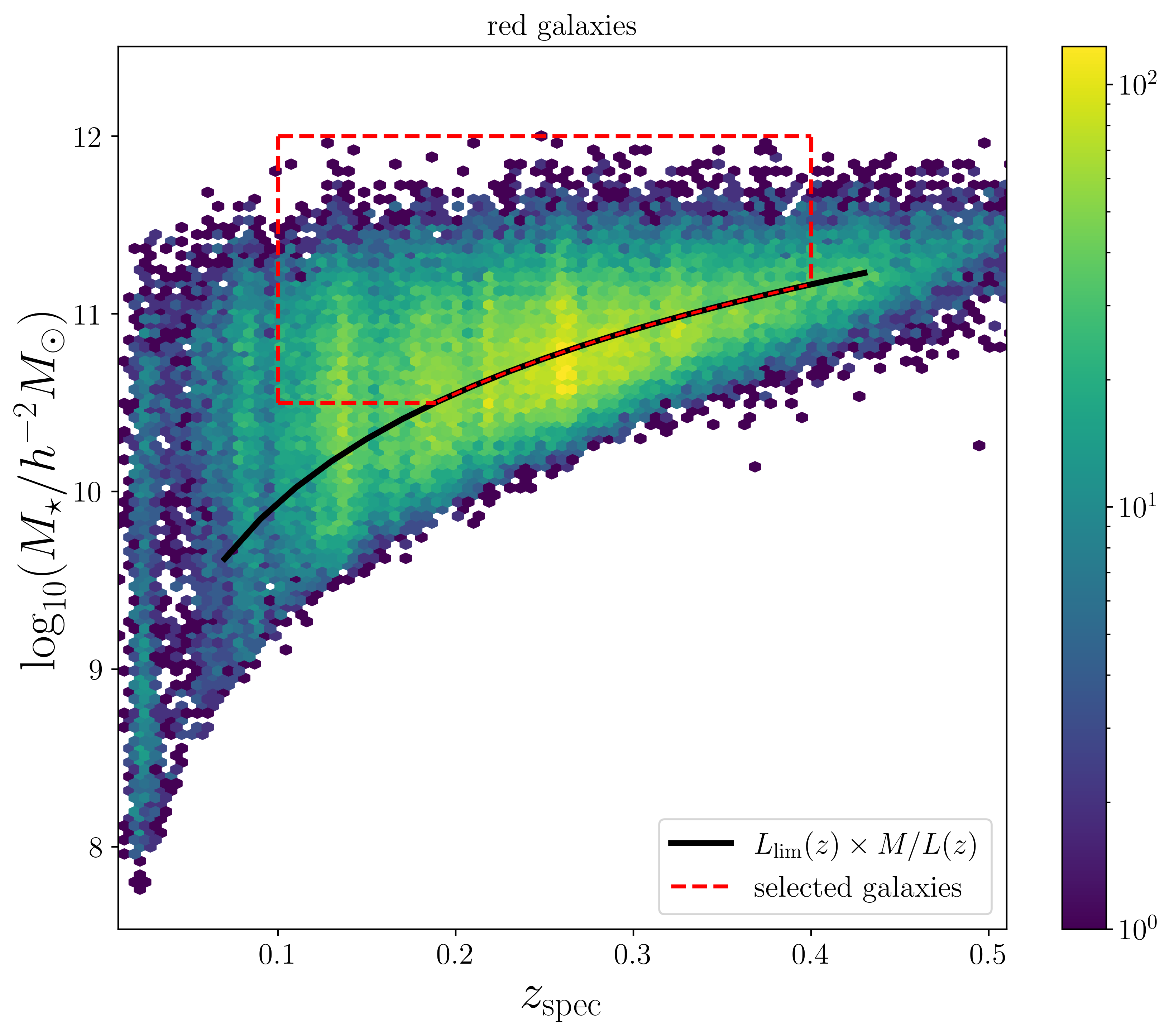}
\caption{Stellar mass vs. spectroscopic distribution. The solid black line shows the derived stellar mass limit. The sample on the left side fulfills the stellar mass completeness criteria. The area enclosed by the red dashed line indicates the sample of galaxies used to build the model in Sect.~\ref{sec:summary_stat}.}
\label{fig:completness_limit}
\end{figure}

With regards to criterion 1, to measure the SMF of the red galaxies in the DESI-EDR, we first have to define the stellar mass limit, which ensures that the sample is complete. We derive the stellar mass limit by measuring the redshift-dependent luminosity limit, which is easily determined using the absolute magnitude in the $r$-band filter  
\begin{equation}
    \log_{10}[L/(h^{-1}L_{\odot})] = 0.4 \times (4.74-M_r) \ ,
\end{equation}
where 4.74 is the magnitude of the Sun in the $r$-band used by the DESI survey. Next, we used the tabulated stellar masses to compute the stellar mass-to-light ratio $M_\star/L(z)$, which is also redshift-dependent. We chose the upper 95$\%$ value for all the redshift bins to be conservative but excluding outliers. Multiplying $M/L(z)$ with $L_\mathrm{lim}(z)$, we then find a stellar-mass limit fitting function motivated by \cite{Zou2019},
\begin{equation}
    \log_{10}[M^\mathrm{lim}_\star/(h^{-1}\mathrm{M}_\odot)] = a \times \log_{10}(b\ z) + c \ ,
    \label{eq:completness}
\end{equation}
where $a = 2.0404 $, $b = 1.8140$ and $c = 11.4499$. In Fig.~\ref{fig:completness_limit}, we show the distribution of stellar mass versus the spectroscopic redshift of the DESI-EDR and the stellar mass completeness function $M^\mathrm{lim}_\star$. To measure the SMF for $10.2 < \log_{10} [M_\star/(h^{-1}\mathrm{M}_\odot)] < 12$ with $\Delta \log_{10} [M_\star/(h^{-1}\mathrm{M}_\odot)] = 0.1$ and $0.1<z<0.4$ from those galaxies that fulfill the stellar mass completeness criteria, we follow \cite{Dvornik2023} and compute for each galaxy its observable volume, where we used the fiducial cosmology \texttt{c000} of \abacussummit\ to measure comoving distances. The SMF is given by binning the stellar masses and summing over all these volumes, weighted by the weight per galaxy provided in the DESI-EDR catalogues for completeness to fibre collisions. Furthermore, we ignore that the DESI samples are missing objects that, although they appear brighter than $19.5$, have an apparent fibre magnitude too faint to properly measure redshifts. This reduces the completeness to $98\%$ \cite{Hahn2023A}. However, even if we multiply the derived SMF with $1/0.8$, which is according to figure 12 of \cite{Hahn2023A}, the worst case, the measured SMF is still inside the allowed prior range (see black dots in Fig.~\ref{eq:SMF}). The SMF uncertainty is estimated by dividing the sample into ten subsamples. To predict SMF, we use Eq.~\ref{eq:SMF}, where the halo mass function $n(\Mh)$ is measured from the 25 phase realizations of the \abacussummit\ to reduce the effect of cosmic variance and the assumed cosmology is the fiducial cosmology \texttt{c000} of \abacussummit. We fixed the cosmology and then allowed only those HOD parameters with a $\chi^2 < 100$, where we used only the diagonal of the covariance matrix. 

\begin{figure}
\includegraphics[width=\linewidth]{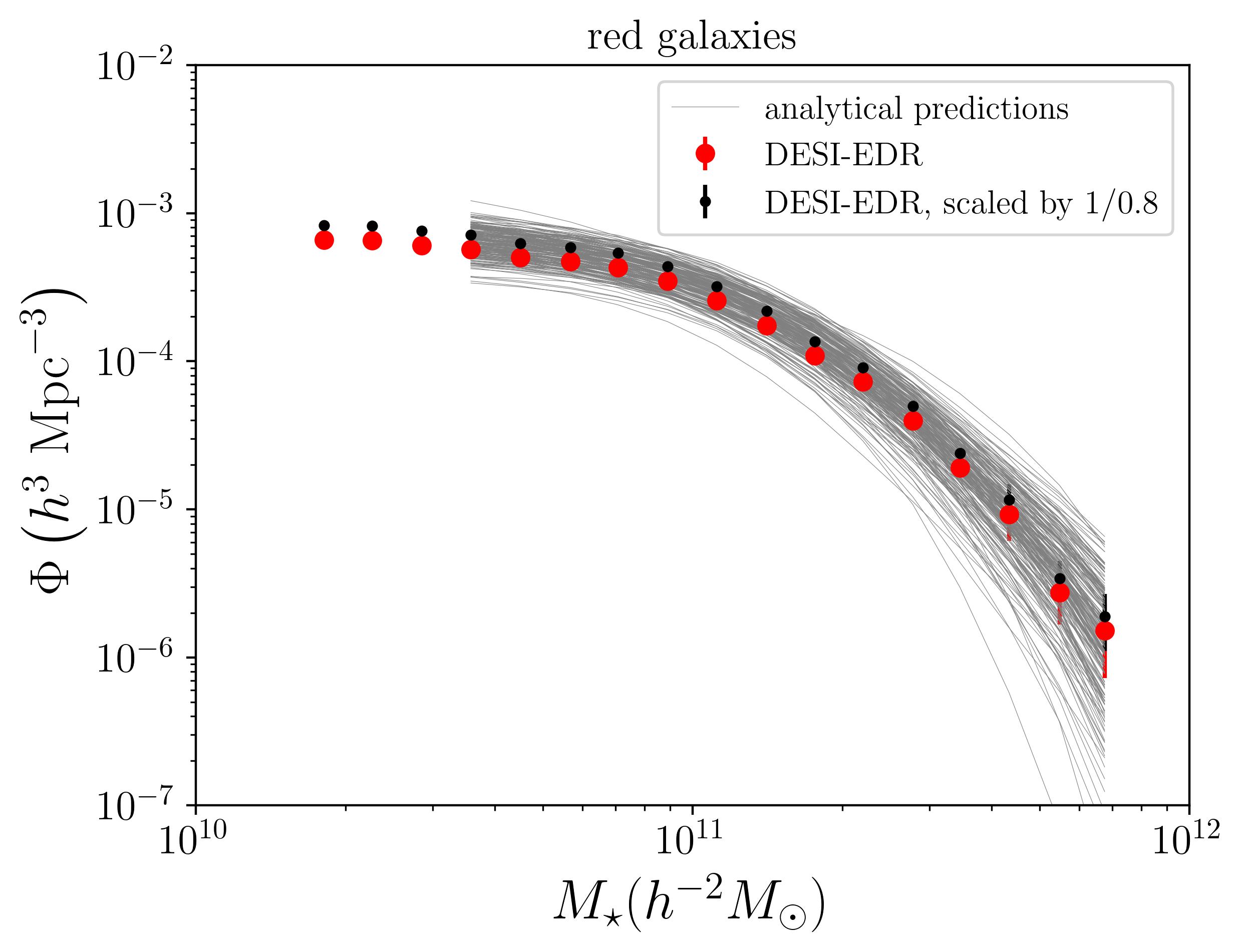}
\caption{SMF of the DESI-EDR described in Sect.~\ref{sec:prior}, and the analytical predictions described in Sect.~\ref{sec:summary_stat} for the first 200 points shown in Fig.~\ref{fig:paramerterspace_CSMF}. The scaling by $1/0.8$ is explained in the main text due to possible incompleteness due to fibre magnitudes being too faint to measure redshifts.}
\label{fig:SMF}
\end{figure}

This threshold of 100 is arbitrarily chosen; however, as shown in Fig.~\ref{fig:SMF}, the resulting SMF predicted with the more sophisticated approach described in Sect.~\ref{sec:summary_stat} spread significantly more than the DESI-EDR measurement suggests. Again, we note that the estimate of the SMF uncertainty, the threshold of 100, and the simplified prediction with a non-varying cosmology are solely used to define the prior range of the HOD parameter space and are not used for the parameter forecasts. We have checked that using a different cosmology has not changed the prior range significantly. 

\begin{figure}
\includegraphics[width=\linewidth]{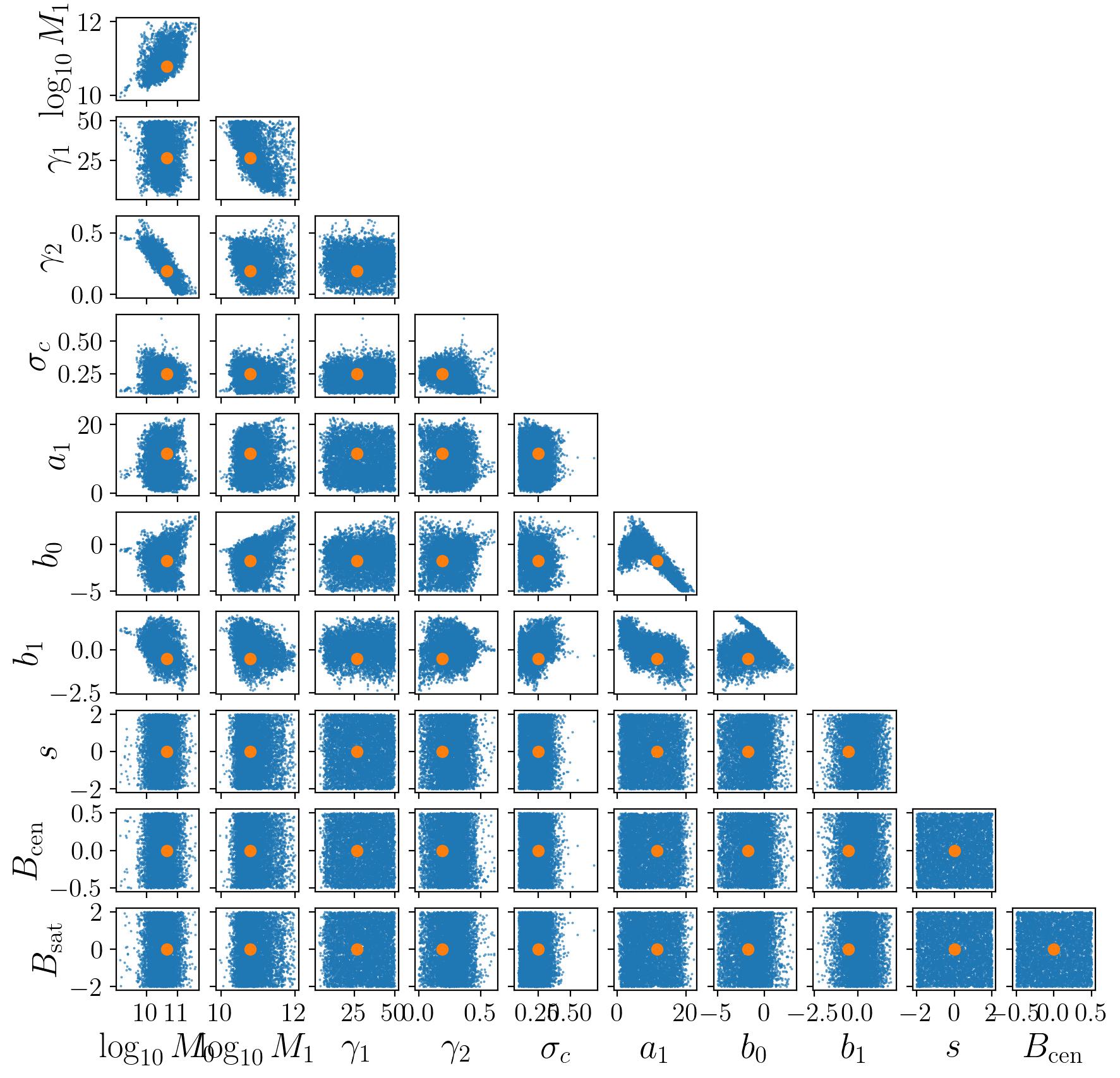}
\caption{Resulting parameter space for the CSMF HOD that fulfills the criteria. The orange dot indicates the reference HOD parameters given in Table~\ref{tab:fiducial}.}
\label{fig:paramerterspace_CSMF}
\end{figure}

\begin{table*}
    \renewcommand{\arraystretch}{1.2}
    \centering
    \caption{Fiducial cosmological and HOD parameters.}
    \begin{tabular}{| l | l | l | l | l | l | l | l | l | l | l | l | l | l | l | l | l |}
        \hline
        Parameter & $\Omega_{\rm b}$ & $\Omega_{\rm cdm}$ & $\sigma_8$ & $n_\mathrm{s}$ & $\nrun$ & $\Neff$ & $w_0$ & $w_a$ & $h$ \\
        \hline
        Value & $0.046$  & $0.233$ & $0.82$  & $0.97$  & $0.0$  & $3.046$  & $-1.0$  & $0.0$ & $0.7$ \\
        \hline
    \end{tabular}
    \vspace{0.1cm}
    \begin{tabular}{| l | l | l | l | l | l | l | l | l | l | l | l | l | l | l | l | l |}
        \hline
        Parameter & $\mathrm{log}_{10} M_{0}$ & $\mathrm{log}_{10} M_1$ & $\sigma_c$ & $\gamma_1$ & $\gamma_2$ & $a_1$ &  $b_0$ & $b_1$ & $B_{\rm cen}$ & $B_{\rm sat}$ & $s$ \\
        \hline
        Value   & $10.6441$   & $10.7807$  & $0.2509$  & $26.7791$  & $0.1900$   & $11.5309$  & $1.7706$ & $-0.5379$ & $0.0$ & $0.0$& $0.0$   \\
        \hline
    \end{tabular}
    \label{tab:fiducial}
\end{table*}

To ensure that the parameters cover the whole valid prior range, we ran an MCMC process with 1000 walkers, where the log-likelihood is always -0.5 if the criteria are fulfilled. If the walkers have been running for long enough, they should cover the whole prior range. From those points in the chain, we randomly selected 6300 (100 per cosmology) combinations as points in parameter space, which we use to train our emulator. The resulting parameter space is shown in Fig.~\ref{fig:paramerterspace_CSMF}, where it is seen how correlated some of the parameters are. The assembly bias and satellite modulation parameter have fixed priors motivated from \cite{Burger2024b}. The lower bounds of $b_0$, $b_1$, and $\gamma_2$ and the upper bounds of $\gamma_1$ are similar to the ones used in \cite{Dvornik2023}. Although these prior restrictions are dominant for some parameters, as we will see in Sect.~\ref{sec:forecast}, they do not affect essential HOD parameters like $M_{0,1}$ or $\sigma_\mathrm{c}$.

\section{Modelling the summary statistics}
\label{sec:summary_stat}
\begin{figure*}
\includegraphics[width=\linewidth]{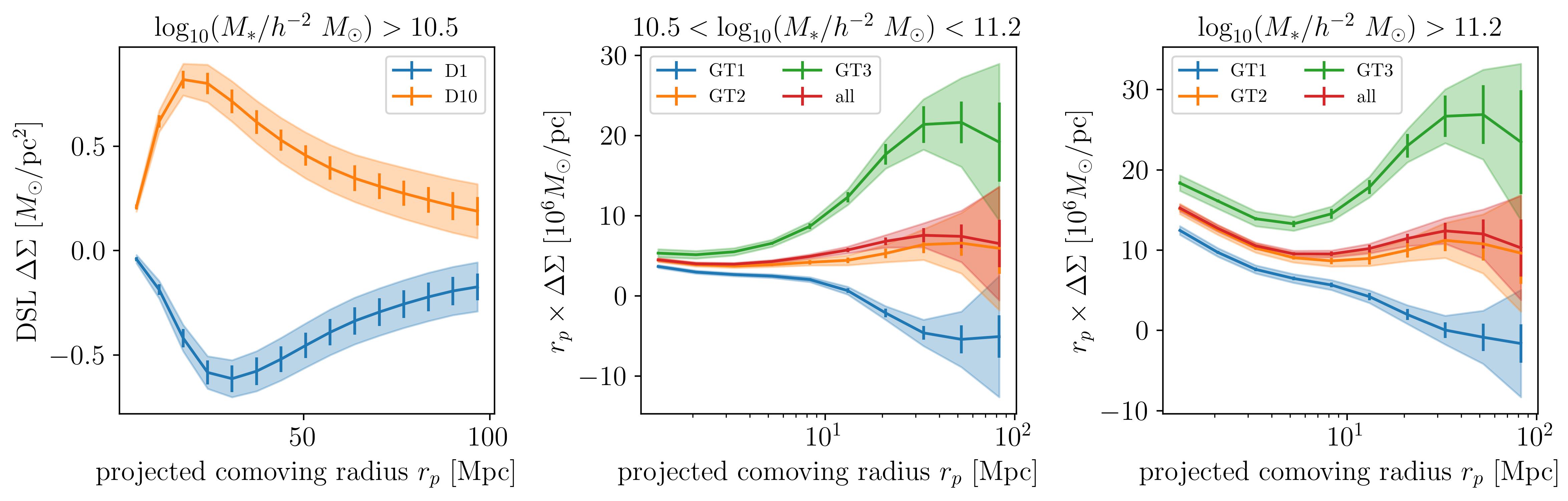}
\caption{Illustration of all ESD data vectors. The left panel shows the ESD around random query points of the highest (D10) and lowest (D1) decile. The middle and right panel shows the ESD around all lens galaxies in red or for the three distinct deciles. The shade region is the expected error estimated from the \citetalias{Takahashi2017} mimicking a possible DESIxUNIONS setup as described in Sect.~\ref{sec:T17}, and the error bars are estimated from the 25-phase realization given in \abacussummit.}
\label{fig:data_vector}
\end{figure*}

In this chapter, we describe our method to model our summary statistics, using \texttt{pyccl}\footnote{\url{https://github.com/LSSTDESC/CCL}} \cite{Chisari2019} to model the SMF and the \abacussummit\ for the ESD measurements. Although we are not comparing our model to real measurements, we still need to assume a fiducial cosmology to measure distances. In this work, we used the \citetalias{Takahashi2017} cosmology, where the relevant parameters are the matter and vacuum energy density with $\Omega_{\rm m}=1-\Omega_\Lambda=0.279$, the baryon density parameter $\Omega_{\rm b}=0.046$, and the dimensionless Hubble constant $h=0.7$

\subsection*{SMF modelling}
To model the SMF for $10.5 < \log_{10} (M_\star/(h^{-1}\mathrm{M}_\odot)) < 12$\footnote{The lower bound is 10.5 instead of 10.2, which was used for the SMF measurements to determine the prior range, since we initially aimed to have volume-limited samples for a better comparison with the literature. However, we dropped that criterion and used all galaxies above $M_\star^\mathrm{lim}$. As long as we treat real data and simulations the same way, this is a fair approach.} with $\Delta \log_{10} (M_\star/(h^{-1}\mathrm{M}_\odot)) = 0.1$ and $0.1<z<0.4$, we use again Eq.~\ref{eq:SMF}, where the halo mass function $n(\Mh)$ is modelled using the function form of \cite{Tinker2010} implemented in \texttt{pyccl}, which changes with cosmology and redshift. To include the redshift dependence, we measured the median redshift of the \citetalias{Takahashi2017} mocks described below over all realizations for each stellar mass bin. Furthermore, to account for the cosmology dependence of potential SMF measurement, we need to modify the SMF prediction $\Phi^\mathrm{model}_\mathrm{X}(M_\star)$ with 
\begin{equation}
    \Phi^\mathrm{fid}_\mathrm{X}(M_\star) = \Phi^\mathrm{model}_\mathrm{X}(M_\star) \frac{\chi^3(z_\mathrm{u},\mathcal{C}^\mathrm{model})-\chi^3(z_\mathrm{l},\mathcal{C}^\mathrm{model})}{\chi^3(z_\mathrm{u},\mathcal{C}^\mathrm{fid})-\chi^3(z_\mathrm{l},\mathcal{C}^\mathrm{fid})} \ ,
\end{equation}
where $\chi$ denotes the comoving distance, $\mathcal{C}^\mathrm{fid}$ the \citetalias{Takahashi2017} cosmology, $\mathcal{C}^\mathrm{model}$ the model cosmology, and $z_\mathrm{l}$ and $z_\mathrm{u}$ the lower and upper redshifts, respectively. While the lower redshift is set $0.1$, the upper redshifts are determined by converting the median stellar mass for each stellar mass bin measured from the \citetalias{Takahashi2017} mocks to limiting redshift values with Eq.~\ref{eq:completness}. 
   
\subsection*{ESD modelling}
To model the ESD, we decided to use two stellar mass bins, with the lower bin being ${10.5 < \log_{10} [M_\star /(h^{-1}\mathrm{M}_\odot)] < 11.2}$ and the larger bin being ${\log_{10} [M_\star /(h^{-1}\mathrm{M}_\odot)] > 11.2}$. The ESD for the two stellar mass bins around the galaxies that fulfill the stellar mass completeness criteria $M_\star^\mathrm{lim}$ are measured in ten logarithmic bins for ${1 < r_\mathrm{p} /(h^{-1}\mathrm{Mpc}) <100}$ are shown in the middle and right panels of Fig.~\ref{fig:data_vector}. Since we assume the \citetalias{Takahashi2017} cosmology to measure distances, we directly incorporate the correction for the cosmology dependence of these measurements. Therefore, we do not have to apply a correction to the ESD measurements as we did for SMF prediction.

\subsection*{Environmental dependence of the ESD model}
In addition to the raw ESD around all the lens galaxies, we aim to measure how the ESD depends on the local galaxy density environments. Motivated by the previous works of \cite{Paillas2023}, we define the three-dimensional environments by smoothing the three-dimensional galaxy distribution, $n_\mathrm{g}$, with a Gaussian filter of $R_G = 8 \ h^{-1}\mathrm{Mpc}$. We note that we used all galaxies with ${\log_{10} [M_\star /(h^{-1}\mathrm{M}_\odot)] > 10.5}$ and $0.1<z<0.3$ that fulfill the completeness criteria to define $n_\mathrm{g}$. The redshift cut of $z<0.3$ results from the fact that above that limit, the number density drops below $\sim 10$ galaxies per sphere of radius $2 \times 8 \ h^{-1}\mathrm{Mpc}$, and therefore, would result in noisy estimates of the environment. To ensure consistency between the ESD measurements, we also performed the same redshift cut for the ESD measurements of all galaxies. To account for masking, we divide the smoothed galaxy distribution, $n_\mathrm{g}^{R_G}$, with a smooth version of random points, $n_\mathrm{r}^{R_G}$, which are distributed in the same volume as the galaxies but several times denser. This corrected density environment, 
\begin{equation}
    \delta_{g}^{R_G} = \frac{n_\mathrm{g}^{R_G}}{n_\mathrm{r}^{R_G}}-1\ ,
\end{equation}
is now used in two distinct ways. 

In the first approach, we follow the previous work of \cite{Burger2024b} by ranking the $\delta_{g}^{R_G}$ with the modification of dividing into ten deciles instead of five quantiles (D1 - D10). Furthermore, we have found that the highest and lowest deciles carry the most constraining power, so we decided to use only the highest and lowest deciles to keep the data vector small. Next, we measure the ESD around random query points inside the deciles D1 and D10. The resulting ESD, which we note as DSL $\Delta \Sigma$, measured in 15 linear bins for ${1 < r_\mathrm{p} /(h^{-1}\mathrm{Mpc}) <100}$ are shown in the left panel of Fig.~\ref{fig:data_vector}. 

In the second approach, we divide the galaxies first into two stellar mass bins ${10.5 < \log_{10} [M_\star /(h^{-1}\mathrm{M}_\odot)] < 11.2}$ and ${\log_{10} [M_\star /(h^{-1}\mathrm{M}_\odot)] > 11.2}$, and assign to each galaxy a $\delta_{g}^{R_G}$ value. Next, we divide the galaxy sample for the two stellar mass bins into three galaxy tertiles (GT) according to their $\delta_{g}^{R_G}$ value. The resulting ESD, which we note as Galaxy-DSL $\Delta \Sigma$, measured in ten logarithmic bins for ${1 < r_\mathrm{p} /(h^{-1}\mathrm{Mpc}) <100}$ are shown in the middle and right panels of Fig.~\ref{fig:data_vector}. As expected, the highest GT results in the ESD with the largest amplitude show that the host haloes of galaxies living in high-density environments are usually more massive.

\begin{figure}
\includegraphics[width=\linewidth]{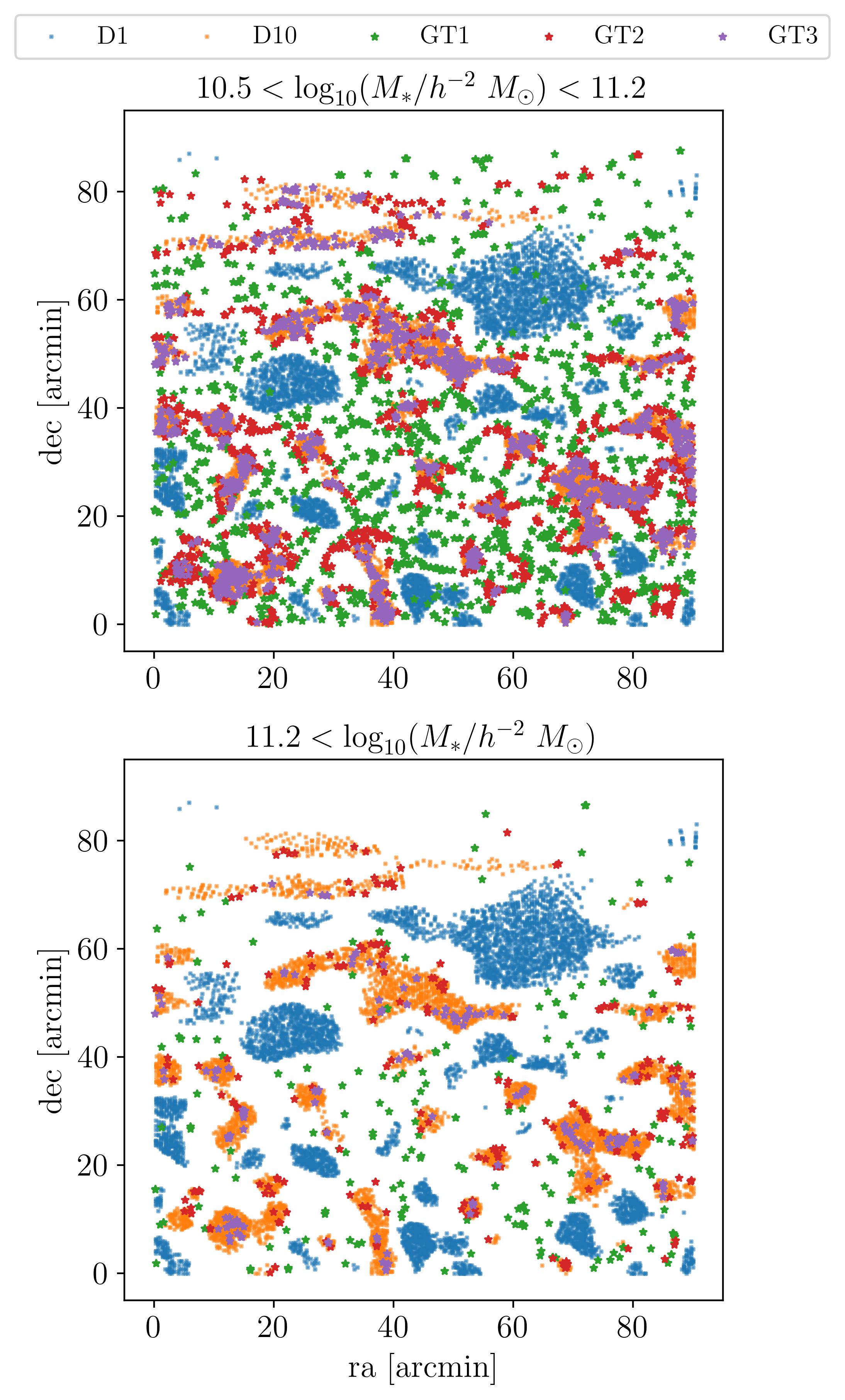}
\caption{Distribution of the D1 and D10 random points, and the distribution of the GT1-3, which depends on the stellar mass bins. This image is a slice with $\chi<310\,\mathrm{Mpc}/h$.}
\label{fig:DSL_galaxy_distribution}
\end{figure}

As for ESD modelling around galaxies, we assumed the \citetalias{Takahashi2017} cosmology for the distance measurements. Therefore, we do not have to apply a correction to the ESD measurements due to their different cosmologies.
Finally, we illustrate in Fig.~\ref{fig:DSL_galaxy_distribution} the distribution of random points of D1 and D10 and the distribution of the galaxies depending on their environment. It is seen that GT2 is just outside D10 while GT3 is more clustered towards the center of D10. The GT1 galaxies are located between D1 and D10 and, therefore, can be identified as field galaxies. 

\section{Estimating a covariance matrix}
\label{sec:T17}
We use as in \cite{Burger2024b} the \citetalias{Takahashi2017} simulations to estimate a covariance matrix of a possible DESIxUNIONS setup. Therefore, we only mention the essential details and deviations from the previous work. The matter and vacuum energy density are $\Omega_{\rm m}=1-\Omega_\Lambda=0.279$, the baryon density parameter $\Omega_{\rm b}=0.046$, the dimensionless Hubble constant $h=0.7$, the power spectrum normalization $\sigma_8=0.82$, and the spectral index $n_{\rm s}=0.97$. 

To measure a DESI-like SMF covariance from the \citetalias{Takahashi2017}, we first populate the haloes with the description in Sect.~\ref{sec:AbacusSummit}, and the reference parameters shown as the orange dot in Fig.~\ref{fig:paramerterspace_CSMF} and given Table~\ref{tab:fiducial}.
Although we know that data release one of DESI will not cover the full UNIONS footprint, we assume it for this work and use the same masking we used for the UNIONS mocks, which we determined from the UNIONS galaxies themselves. Therefore, the parameter forecasts in this paper are over-optimistic for at least the first data releases of DESI. Given the stellar mass limit in Eq.~\ref{eq:completness}, we next measure the SMF for $10.5 < \log_{10} (M_\star/(h^{-1}\mathrm{M}_\odot)) < 12$ with $\Delta \log_{10} (M_\star/(h^{-1}\mathrm{M}_\odot)) = 0.1$ and $0.1<z<0.4$. 

The next step is to measure the $\Delta \Sigma$'s. Since the ESD is independent of the source redshift distribution, we selected a single redshift at $z=0.64$, which is behind the lens distribution. We use the highest available resolution, which has a pixel resolution of $A_\mathrm{pix}=0.18\,\mathrm{arcmin}^2$ ($\texttt{nside}=8192$), or a side length of $98.8\,h^{-1}\mathrm{kpc}$ at a redshift of $z=0.4$ given the \citetalias{Takahashi2017} cosmology. As in \cite{Burger2024b}, we create galaxy shear catalogues by extracting the shear information at each galaxy position from the preliminary UNIONS catalogue (v1.3 ShapePipe), covering roughly a footprint of $2800\,\mathrm{deg}^2$. We add shape noise contribution, $\epsilon^{\mathrm{s}}$, to the reduced shear $g = \gamma/(1-\kappa)$ as follows \citep{Seitz1997}:
\begin{equation}
\epsilon^{\mathrm{obs}} = \frac{{ \epsilon}^{\mathrm{s}}+{ g}}{1+{ \epsilon}^\mathrm{s}{g}^*} \, ,
\label{eq:ebos}
\end{equation}
where asterisk `$*$' indicates complex conjugation, and $\epsilon^{\mathrm{s}}$ is determined from randomly rotating the observed ellipticities of the preliminary UNIONS catalogue. 

\begin{figure}
\includegraphics[width=\linewidth]{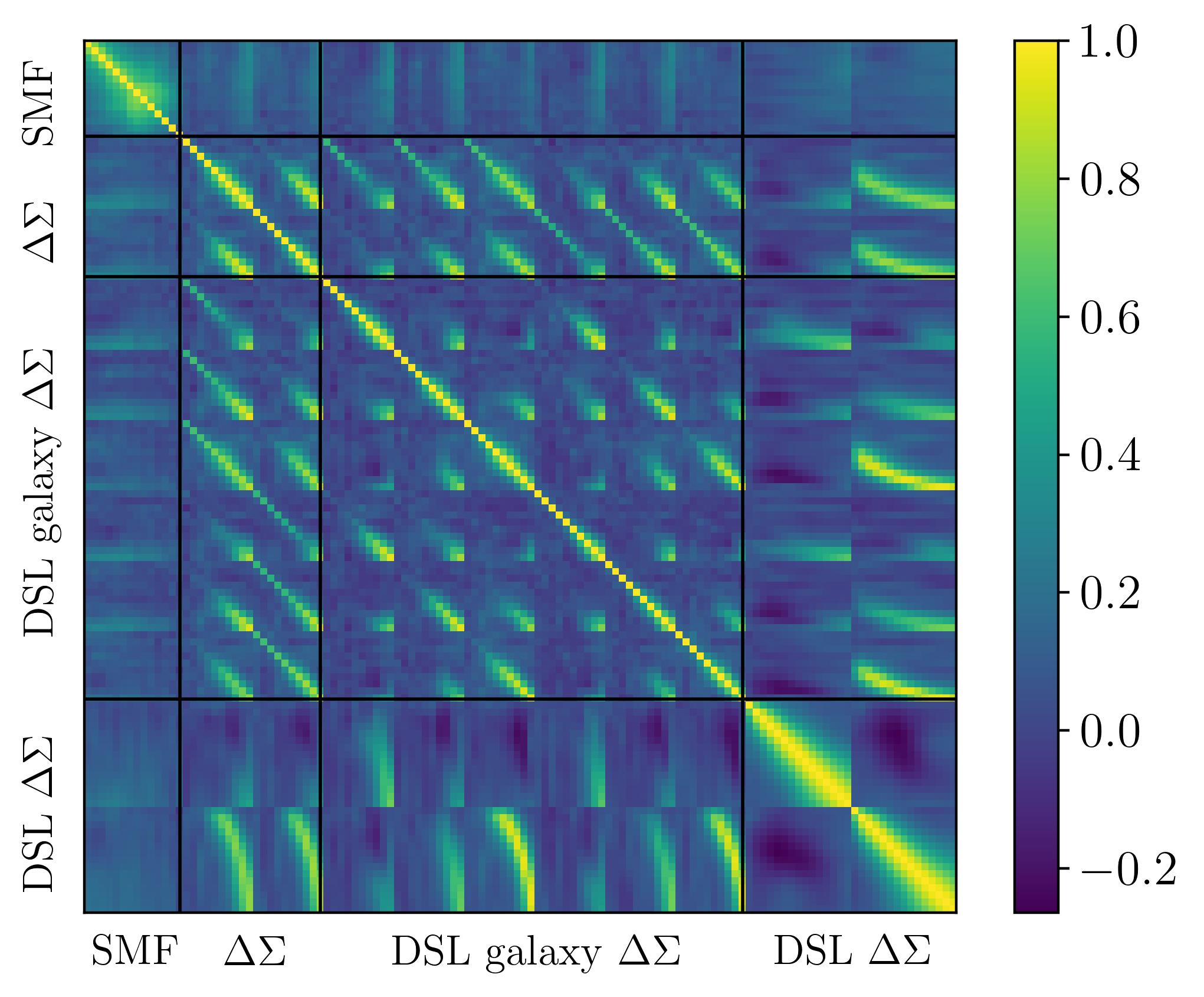}
\caption{Correlation matrix of the whole data vector.}
\label{fig:correlation_matrix}
\end{figure}

By rotating the galaxy positions and ellipticities seven times by $50^\circ$ along the lines of constant declination, we created 756 pseudo-independent mock catalogues from which the covariance is measured. The resulting correlation matrix is shown in Fig.~\ref{fig:correlation_matrix}. Not surprisingly, there is almost no correlation between the SMF and the ESD. 

\section{Building the emulator}
\label{sec:emulator}
To build the emulator, we closely follow the procedure of \cite{Burger2024b}, with the exception that we are not averaging multiple noise seeds of the HOD, as the sample size for the used BGS mocks is much greater than the LOWZ sample used in \cite{Burger2024b}, and the measurement of the ESD is more time-consuming. However, to reduce the effect of cosmic variance, we average over the $n_\mathrm{p}=25$ phases. For a given HOD $\texttt{h}$, the corrected model is given by
\begin{equation}
    \boldsymbol{m}^\mathrm{cor}(\Theta;\texttt{h}) = \frac{\boldsymbol{m}^\texttt{ph00}(\Theta)}{\boldsymbol{m}^\texttt{ph00}(\texttt{c000};\texttt{h})}  \frac{1}{n_\mathrm{p}}\sum_{p=1}^{n_\mathrm{p}}  \boldsymbol{m}^p(\texttt{c000};\texttt{h}) \,  \, ,
    \label{eq:model2}
\end{equation}
where $\Theta$ indicates a set of parameters, $\texttt{ph00}$ the phase for which we have multiple cosmology realizations, and $\texttt{c000}$ the fiducial cosmology for which we have $n_\mathrm{p}=25$ phase realizations. Since the corrected model depends on the reference HOD parameters \texttt{h}, we average over the $n_\mathrm{\texttt{h}}=20$ reference HOD parameters, which results in the final model
\begin{equation}
    \boldsymbol{m}(\Theta) = \frac{1}{n_\mathrm{\texttt{h}}} \sum_{\texttt{h}=1}^{n_\mathrm{\texttt{h}}} \bar{\boldsymbol{m}}(\Theta,\texttt{h}) \, .
    \label{eq:model4}
\end{equation}
Finally, we applied, as in \cite{Burger2024b}, a principal component analysis (PCA) as a final noise reduction \cite{EE2021,Arico2021}, where we used only the first four principal components for each measurement. 

\begin{figure}
\includegraphics[width=\linewidth]{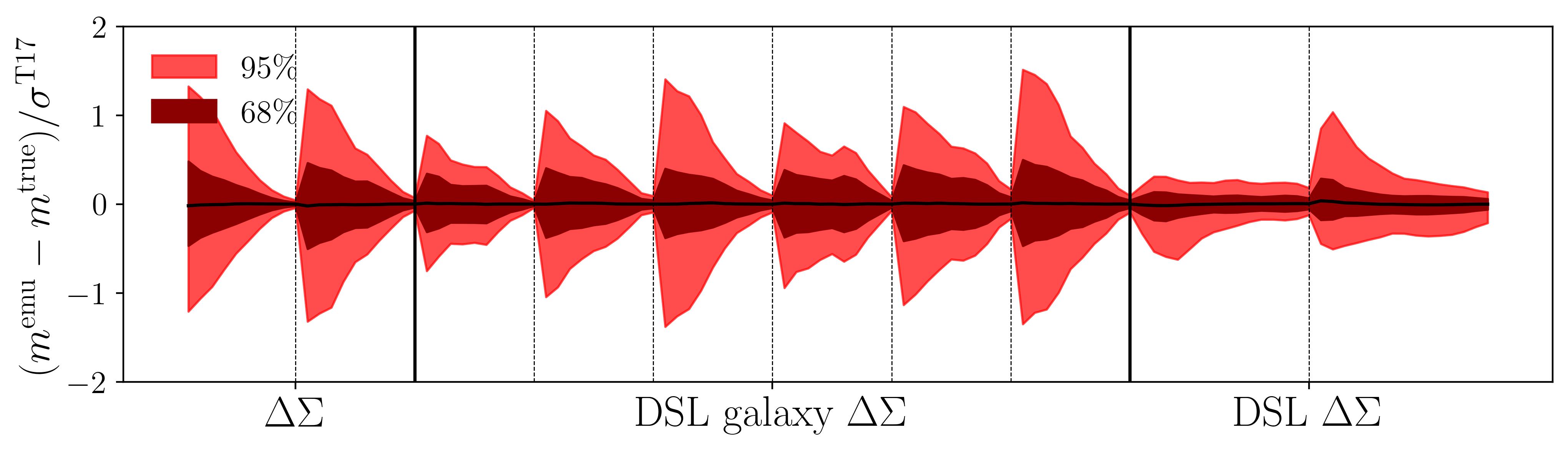}
\caption{Residual emulator accuracy compared to the \citetalias{Takahashi2017} covariance matrix. The error is scaled according to Eq.~\eqref{eq:percival_correction}.}
\label{fig:emulator_acc}
\end{figure}

\begin{figure}
\includegraphics[width=\linewidth]{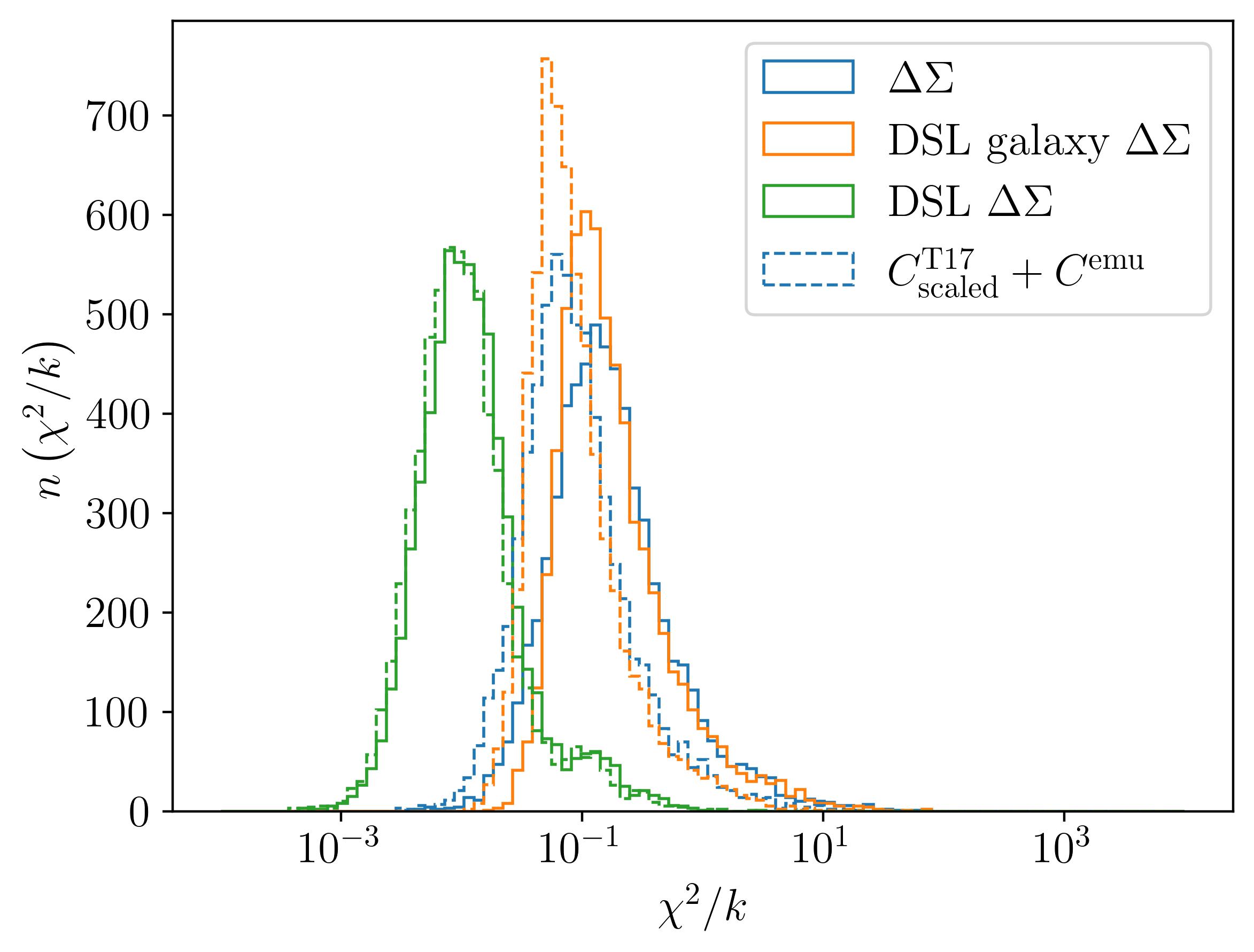}
\caption{Distributions of reduced $\chi^2$ between true and emulated measurements for all 6300 measurements.}
\label{fig:chi2}
\end{figure}

To quantify the error on the emulator, we perform the leave-one-out method to quantify the overall error in the model. As in \cite{Burger2024b}, we successively leave out one of the 63 cosmologies and train on the other 62, which gives us 6300 emulator predictions on parameters that the trained emulator has not seen during the training. The emulator accuracy with respect to the \citetalias{Takahashi2017} covariance matrix is shown in Fig.~\ref{fig:emulator_acc}. However, since the residual plot might be misleading, we measure a $\chi^2$ between the true and emulated measurements for all 6300 measurements. As shown in Fig.~\ref{fig:chi2}, the resulting $\chi^2$ is mostly smaller than one, indicating that the emulator is accurate enough for the DESIxUNIONS uncertainty. 

As the dashed lines in Fig.~\ref{fig:chi2}, we show the improvement on the $\chi^2$ distribution if we adding the emulator error to the \citetalias{Takahashi2017} covariance 
\begin{equation}
    \label{eq:emu_cov}
    C = C^\mathrm{T17} + C^\mathrm{emu} \,,
\end{equation}
where we estimate $C^\mathrm{emu}$ using the scatter in quantity $\Delta m = m^\mathrm{emu}-m^\mathrm{true}$. To account for the fact that both $C^\mathrm{T17}$ and $C$ are measured from noisy realisations \cite{Percival2021}, we correct them by 
\begin{equation}
\label{eq:percival_correction}
    C' = C  \frac{(n_\mathrm{r}-1)[1+B(n_\mathrm{d}-n_\Theta)]}{n_\mathrm{r}-n_\mathrm{d}+n_\Theta -1}
\end{equation}
where
\begin{equation}
    B = \frac{n_\mathrm{r}-n_\mathrm{d}-2}{(n_\mathrm{r}-n_\mathrm{d}-1)(n_\mathrm{r}-n_\mathrm{d}-4)} \, .
    \label{eq:B}
\end{equation}
and $n_\mathrm{r}=756$ is the number of realizations, $n_{\Theta}=12$ the number of free parameters, and $n_{\rm d}$ the rank of the covariance matrix. Although $n_\mathrm{r}$ for $C^\mathrm{emu}$ is not clearly defined, we used the same correction for $C$ as we used for $C^\mathrm{T17}$. 

\section{Forecast}
\label{sec:forecast}

With the modelling and covariance matrix described in the previous chapters, we focus now on a parameter constraint forecast. Since the estimated covariance matrix $C$ is a random variable itself, we follow \cite{Percival2021} to compute the likelihood. Given a data vector $\boldsymbol{d}$ and covariance matrix $C$ of rank $n_\mathrm{d}$ measured from $n_\mathrm{r}$ realizations, the posterior distribution of a model vector $\boldsymbol{m}(\Theta)$ that depends on $n_{\Theta}=12$ parameters is
\begin{equation}
\boldsymbol{P}\left(\boldsymbol{m}(\Theta)|\boldsymbol{d},C\right) \propto |C|^{-\frac{1}{2}} \left( 1 + \frac{\chi^2}{n_{\rm r}-1}\right)^{-m/2}\, ,
    \label{eq:t_distribution}
\end{equation}
where
\begin{equation}
\chi^2 =  \left[\boldsymbol{m}(\Theta)-\boldsymbol{d}\right]^{\rm T} C^{-1} \left[\boldsymbol{m}(\Theta)-\boldsymbol{d}\right] \, .
\label{eq:chi2}
\end{equation}
The power-law index $m$ is 
\begin{equation}
    m = n_\Theta+2+\frac{n_\mathrm{r}-1+B(n_\mathrm{d}-n_\Theta)}{1+B(n_\mathrm{d}-n_\Theta)} \, ,
    \label{eq:m_power}
\end{equation}
and $B$ is defined in Eq.~\eqref{eq:B}.
Although we varied 15 parameters ($\omega_\mathrm{cdm}$, $\omega_\mathrm{b}$, $\sigma_8$, $n_\mathrm{s}$ and the the 11 HOD parameters described in Sec. \ref{sec:AbacusSummit}) during our MCMC, only 12 were not constrained by the priors ($\omega_\mathrm{b}$, $n_\mathrm{s}$ and $s$ were bound by priors). Therefore, we set $n_\Theta=12$. The effective number of parameters should be determined more sophisticatedly in a future analysis. We note that this approach is approximately the same as using a Gaussian approximation in Eq.~\eqref{eq:t_distribution} and using Eq.~\eqref{eq:percival_correction} to correct for the fact that the covariance is measured from simulations. To account for the inaccuracies in our emulator for the ESD\footnote{Inaccuracies in the emulator of the SMF are negligible.}, we decided to use Eq.~\eqref{eq:emu_cov} and treat the power-law index $m$ in Eq.~\eqref{eq:m_power} as it would have measured from $n_\mathrm{r}=756$. Alternatively, we could have used that 
\begin{align}
    C^{-1} &= \left(C^\mathrm{T17} + C^\mathrm{emu} \right)^{-1} \\
    & =  \left(C^\mathrm{T17}\right)^{-1} + \left(C^\mathrm{T17}\right)^{-1}C^\mathrm{emu} \left(C^\mathrm{T17} + C^\mathrm{emu}\right)^{-1} \nonumber \\
    & \approx  \left(C^\mathrm{T17}\right)^{-1} + \left(C^\mathrm{T17}\right)^{-1}C^\mathrm{emu} \left(C^\mathrm{T17}\right)^{-1} \, , \nonumber
\end{align}
where in the last step we assumed that $C^\mathrm{emu} \ll C^\mathrm{T17}$. This would, in principle, allow us to apply the correction in Eq.~\ref{eq:percival_correction} only to the $C^\mathrm{T17}$ and leaving $C^\mathrm{emu}$ untouched, which should improve the constraining power. However, $C^\mathrm{emu} \approx C^\mathrm{T17}$ for some scales, which did not allow us to apply that method. A completely alternative approach to incorporate the error of the emulator would be to use a Bayesian neural network, which finds a likelihood distribution for the parameters for each hidden node of a neural network instead of its best values. This allows us to get a sample of predictions for each input parameter, from which the mean and its uncertainty can be measured. We have tried this but did not achieve a similar accuracy as with \texttt{CosmoPower}\footnote{\url{https://alessiospuriomancini.github.io/cosmopower/}}, so we left it for future analysis to explore it in more detail. 

\begin{figure*}
\includegraphics[width=\linewidth]{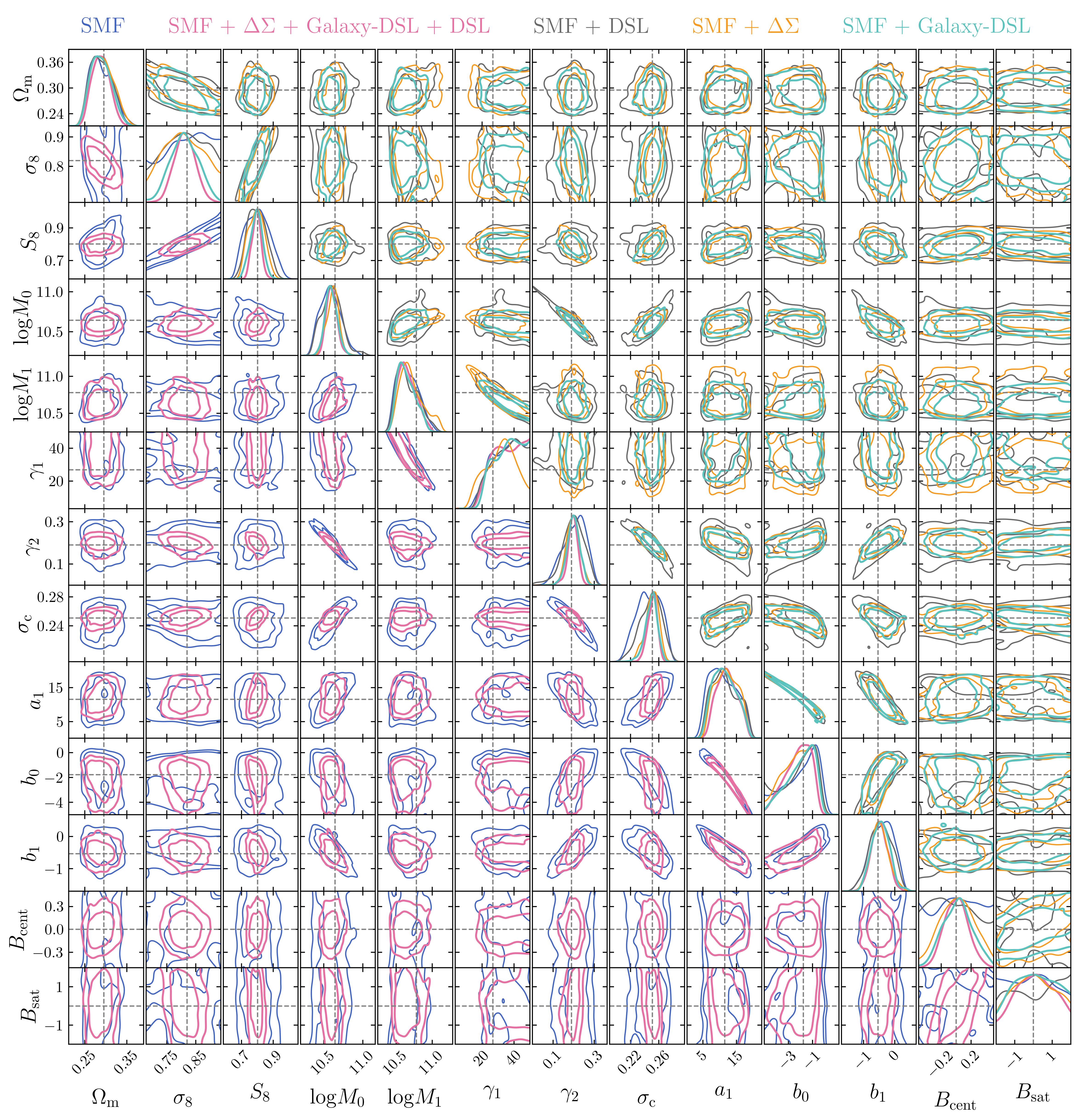}
\caption{Posterior distribution using the emulator itself as a reference predicted at the parameters given in Table \ref{tab:fiducial} and the emulator for the model for all the ESD statistics combined with SMF.}
\label{fig:MCMC}
\end{figure*}

In Fig.~\ref{fig:MCMC}, we show the resulting forecast, where we varied the cosmological parameters $\omega_\mathrm{cdm}$, $\omega_\mathrm{b}$, $n_\mathrm{s}$ and $\sigma_8$. We show only $\Omega_\mathrm{m}$, $\sigma_8$ and $S_8 = \sigma_8 \sqrt{\Omega_\mathrm{m}/0.3}$ as these are the interesting cosmological parameters that we can constrain. The reference data vector is derived using the emulator evaluated at the parameter combination given in Table~\ref{tab:fiducial}. The HOD parameters follow the prior restriction described in Sect.~\ref{sec:prior}. For the MCMC run, we use the \texttt{emcee}\footnote{\url{https://emcee.readthedocs.io/en/stable}} software \cite{emcee2013} with 100 walkers and 10000 steps per walker. We have tested if the posteriors change if we modify these numbers, and the results were stable. Considering each of the three statistics individually, always combined with the SMF, we see that the DSL provides the lowest constraining power, which is expected given the smaller signal-to-noise ratio seen in Fig.~\ref{fig:data_vector}, which in turn results from the fact that the signal at small scales is depleted by measuring it around random points. When using the Galaxy-DSL statistic, we have the highest gain in constraining power, especially in the $\sigma_8$ parameter and, therefore, the $S_8$ parameter. Except for the $B_\mathrm{cen}$ parameter, all HOD parameters are already well constrained by ESD around all galaxies, and the Galaxy-DSL is not bringing significantly more constraining power. However, the environmental ESD significantly improves the constraints for $ B_\mathrm{cen}$.

Overall, we see that the combination of all statistics is powerful in constraining cosmological and HOD parameters and is, therefore, a compelling statistic for future surveys like DESI and UNIONS.

\section{Conclusion}
\label{sec:conclusion}

This paper presents a simulation-based modelling of stellar mass function (SMF) and excess-surface mass density (ESD) for an environmentally dependent sample. 
We use a conditional stellar mass function to model the galaxy-halo connection. Given the halo mass, this model predicts the number of satellites and central and stellar masses for these galaxies. 
After splitting these galaxies into two stellar mass bins, we measure the ESD using the shear information of background galaxies. Furthermore, we use the full sample to measure the SMF and define density environments. These density environments are then used in two distinct methods. The first method, density split lensing (DSL), divides the density environments into ten deciles. Using random points from the highest and lowest deciles, we measure the ESD for those random points. We call this the DSL statistic. The second method divides the sample into two stellar mass bins and then assigns a density value to each galaxy using the density environments computed from the full sample. The galaxies are divided into three tertiles for each stellar mass bin and then used to measure the ESD. We call this the Galaxy-DSL statistic.  

We use the \texttt{AbacusSummit} simulations to build a model for these statistics. Using these simulations, we build an emulator and find that the emulator accuracy is sufficient for a DESIxUNIONS analysis. To account for the remaining noise in the emulator, we derive an emulator covariance matrix using the leave-one-out method. For the parameter forecast analysis, we add this emulator noise to the DESIxUNIONS-like covariance matrix measured from \citetalias{Takahashi2017} simulations.

The methodology presented in this paper is designed for DESIxUNIONS but is also applicable to Stage IV surveys such as Euclid \cite{Laureijs:2011}. While it may not be feasible for Euclid’s first data release (DR1), because of the limited overlap on the sky between Euclid DR1 and spectroscopic surveys, the shear measurements of future Euclid data releases could be combined with low-redshift spectroscopic surveys like DESI or BOSS. For Euclid’s first data release, a similar analysis could be conducted using the internally measured spectroscopic data from NISP \cite{NISP2024}. However, as NISP primarily measures redshifts at  $z > 0.8$ , this introduces additional challenges. Specifically, the usable volume of the \abacussummit\ simulations diminishes significantly for $z > 0.8$, leading to increased noise in the emulator.

We find that environmental-based ESD is beneficial in constraining cosmological and HOD parameters. The satellite modulation parameter $s$ is the exception, which is only measurable with smaller scales. As expected, the cosmological parameters $\omega_b$ and $n_\mathrm{s}$ are also unconstrained with our statistic. We can report constraints on all remaining parameters. 

Overall, our summary statistics provide a reliable and powerful tool to constrain cosmological parameters and understand galaxy evolution regarding assembly bias that depends on the environment. Furthermore, we conclude that the \texttt{AbacusSummit} simulations are reasonably consistent with numerical simulations like the \citetalias{Takahashi2017} simulations. Therefore, \texttt{AbacusSummit} simulations provide an excellent opportunity to validate results from analytical models in the literature, derive better constraining powers, and shed light on new aspects of the galaxy halo connection. We look forward to applying our emulator to DESIxUNIONS data. 


\begin{acknowledgements}

We thank Enrique Paillas, Hanyu Zhang, and Marco Bonici for fruitful comments regarding \abacussummit\ and other analysis details. We want to thank Mike Jarvis for developing and maintaining \texttt{treecorr}, and Alessio Spurio Mancini for developing \texttt{CosmoPower} emulator, which provided a significant part of our model. Furthermore, we thank the authors of \texttt{AbacusSummit} and \texttt{AbacusHOD} for providing and maintaining these excellent simulations and code.

We thank the UNIONS team, for providing us with preliminary shear catalogues and redshift distributions, which allowed us to perform a realistic parameter forecast.

We acknowledge financial support from the Canadian Space Agency (Grant 23EXPROSS1), the Waterloo Centre for Astrophysics and the National Science and Engineering Research Council Discovery Grants program.

This research was enabled in part by support provided by Calcul Quebec\footnote{\url{https://docs.alliancecan.ca/wiki/Narval}} and Compute Ontario\footnote{\url{https://docs.alliancecan.ca/wiki/Graham}} and the Digital Research Alliance of Canada\footnote{\url{alliancecan.ca}}. This research used resources from the National Energy Research Scientific Computing Center, supported by the Office of Science of the U.S. Department of Energy under Contract No. DE-AC02-05CH11231.
\end{acknowledgements}

\appendix

\section{Pseudo simulation-based inference}
In this section, we investigated an alternative sampling strategy to the standard MCMC approach, which is based on the method of simulation-based inference (SBI). The general idea of SBI is to perform a parameter forecast if the likelihood is not well approximated with a Student's $t$ or a Gaussian distribution \cite{Gatti2024,vWK2024}. In other words, we do not approximate the likelihood as given in Eq. \eqref{eq:t_distribution} and then use an MCMC approach to sample this likelihood, but instead use a machine learning approach to learn the likelihood or posterior directly. In these works, the measured summary statistics $\mathbf{t}$ vary with parameters $\Theta$, where the $\Theta$ include cosmological, astrophysical and nuisance parameters. Moreover, one needs to incorporate the noise of the data into the measured $\mathbf{t}$ (i.e. shape and shape noise). The posterior is then estimated by minimizing the Kullback-Leibler divergence \cite{kullback1951information} between the parametric density estimator $p(\Theta|\mathbf{t}; \mathbf{w})$, where $\mathbf{w}$ are the weights of the neural network, and the target distribution $p^*(\Theta|\mathbf{t})$:
\begin{equation}
    D_\mathrm{KL}(p^*|p) = \int p^*(\Theta|\mathbf{t}) \ \ln\left(\frac{p(\Theta|\mathbf{t}; \mathbf{w})}{p^*(\Theta|\mathbf{t})}\right) \ \mathrm{d} \mathbf{t} \ .
\end{equation}
Since the target density is unknown, and only  $\{\mathbf{t}, \Theta\}$ are known, the negative log loss function is computed by
\begin{equation}
    - \ln U(\mathbf{w}|\mathbf{t}, \Theta) = - \sum_{i=1}^{N_\mathrm{samples}} \ln [p(\Theta_i|\mathbf{t}_i; \mathbf{w})] \, ,
\end{equation}
where $N_\mathrm{samples}$ is the number of measured summary statistics. 

Therefore, to use this approach, we must create measurements that include the data's noise. As this is not possible using the \abacussummit\, we decided to create the summary statistics $\mathbf{t}$ as follows:
\begin{enumerate}
    \item Use the emulator to predict a noise-free $\mathbf{t}$;
    \item draw a multivariate random vector $G(0,C^\mathrm{T17}+C^\mathrm{emu})$, and add it to $\mathbf{t}$.
\end{enumerate}
We call this a pseudo-SBI, as we add the noise to the data using a multivariate Gaussian distribution instead of directly incorporating it into the simulations. 

To derive the posteriors we use the \texttt{sbi}\footnote{\url{https://github.com/sbi-dev/sbi}} package \cite{sbi2020}, which is the successor of \texttt{pydelfi} \cite{Alsing2019}. In particular, we are using the neural posterior estimator version A (\texttt{NPE-A}), which uses Bayesian Conditional Density Estimation \cite{Papamakarios2016} to estimate the posterior, which in turn uses a mixture density network \cite{370fbeadb5584ba9ab2938431fc4f140}. We provided a flat prior during the training and removed all points of the chain after evaluating $p(\Theta|\mathbf{t}; \mathbf{w})$ at $\mathbf{t}=\mathbf{t}_0$, that fell outside our non-flat prior given in Fig.~\ref{fig:paramerterspace_CSMF}. The reference $\mathbf{t}_0$ is the noise-free prediction of the emulator at the parameters given in Table \ref{tab:fiducial}.

\begin{figure*}
\includegraphics[width=\linewidth]{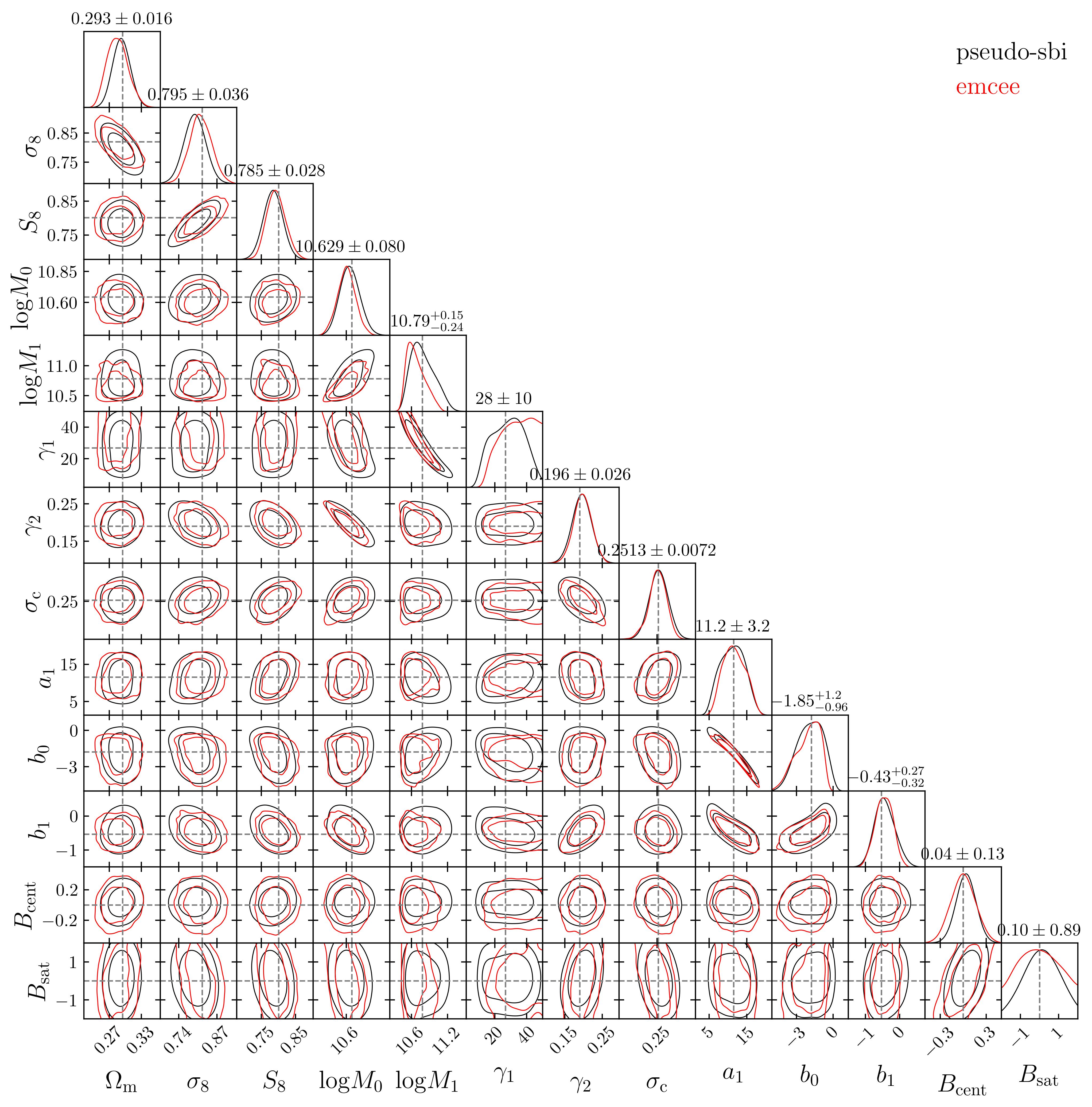}
\caption{Posterior distribution using the pseudo-SBI approach and its comparison to the standard MCMC approach.}
\label{fig:MCMC_sbi}
\end{figure*}

In Fig.~\ref{fig:MCMC_sbi}, we compare the MCMC approach to the pseudo-SBI. Although we added Gaussian noise to the noise-free $\mathbf{t}$, we do not expect that both approaches result in identical posteriors. One reason is that standard MCMC and SBI are approximating the posterior differently. Another reason is that SBI seems to have smoother posteriors, which could mean that the SBI is smoothing over the remaining noise in the emulator. Therefore, we are unsurprised that the pseudo-SBI posterior results in a slightly enlarged posterior distribution. Although the pseudo-SBI gives slightly larger constraints, we want to emphasize the match, especially in the cosmological parameters. Furthermore, it is worth mentioning that the pseudo-SBI produces converged posteriors roughly five times faster than \texttt{emcee}, depending on the number of walkers and steps of the MCMC process and the measurements to train to the neural network of the SBI.

\section{T17 analysis}
\label{Sec:T17_analysis}

\begin{figure*}
\includegraphics[width=\linewidth]{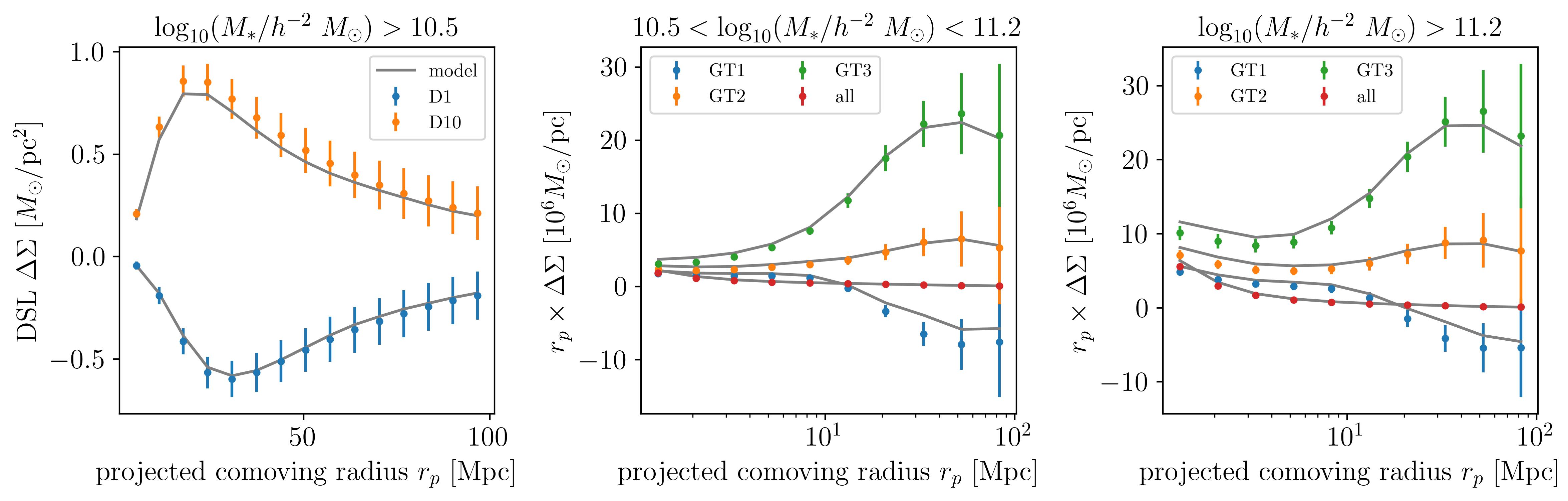}
\caption{Emulator predictions and \citetalias{Takahashi2017} measurements.}
\label{fig:ESD_T17}
\end{figure*}

\begin{figure}
\includegraphics[width=\linewidth]{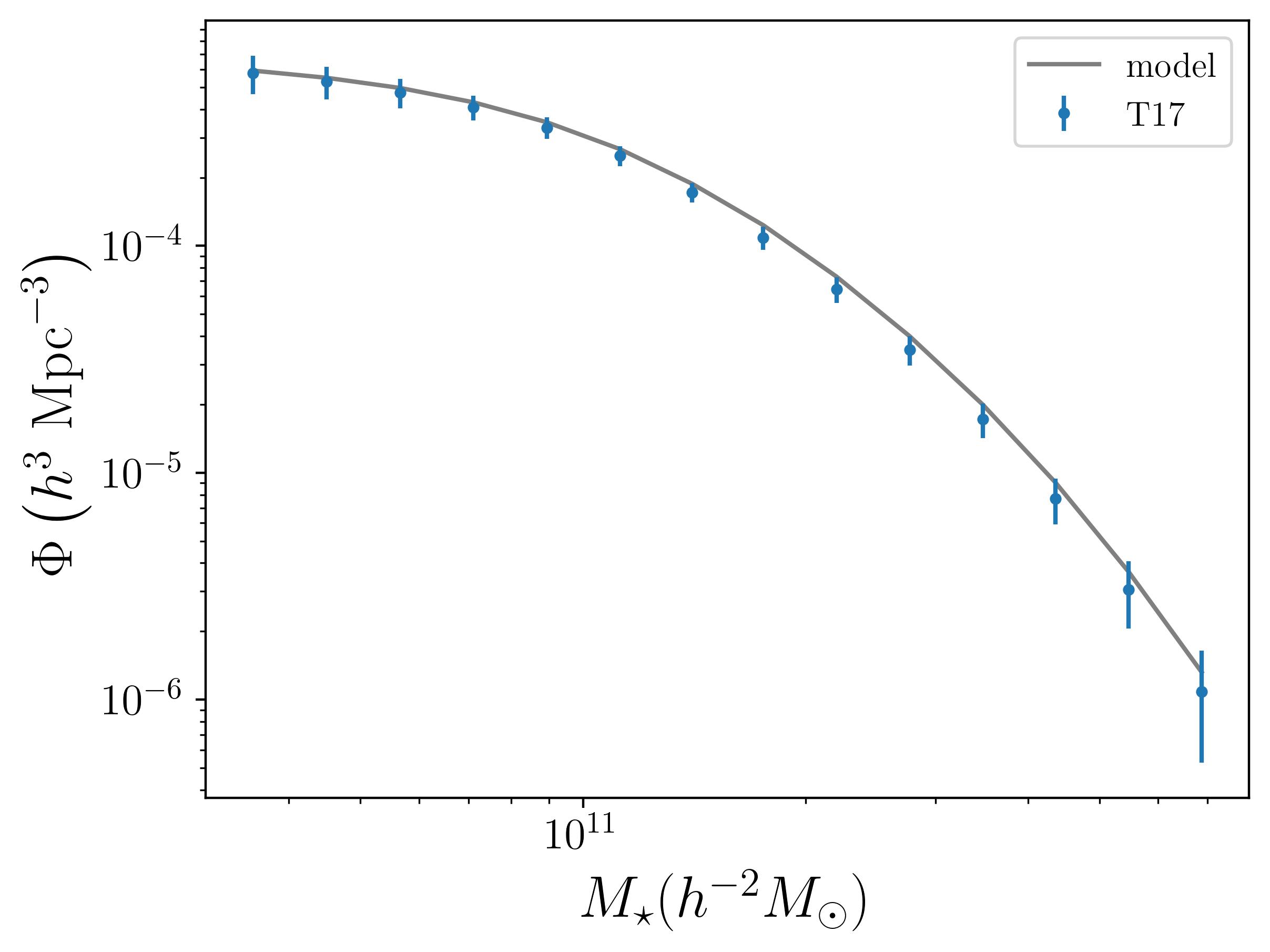}
\caption{SMF predictions using \texttt{pyccl} and the \citetalias{Takahashi2017} measurements. The error bars are scaled by ten to make them visible.}
\label{fig:SMF_T17}
\end{figure}

Although we could have used the \citetalias{Takahashi2017} simulations as our reference data vector to check if our model gives unbiased results when compared with a different suite of simulations, we decided not to give this a huge attention. The reason is that \citetalias{Takahashi2017} used different mass resolutions and halo finder than the \abacussummit, have not included a massive neutrino, and we need to use an NFW profile to populate satellites instead of using the particles. However, we emphasize that the measured \citetalias{Takahashi2017} mocks have the same order of magnitude as the \abacussummit\ predictions as shown in Fig.~\ref{fig:ESD_T17} and \ref{fig:SMF_T17}, indicating that the covariance matrix is realistic. 

\begin{figure*}
\includegraphics[width=\linewidth]{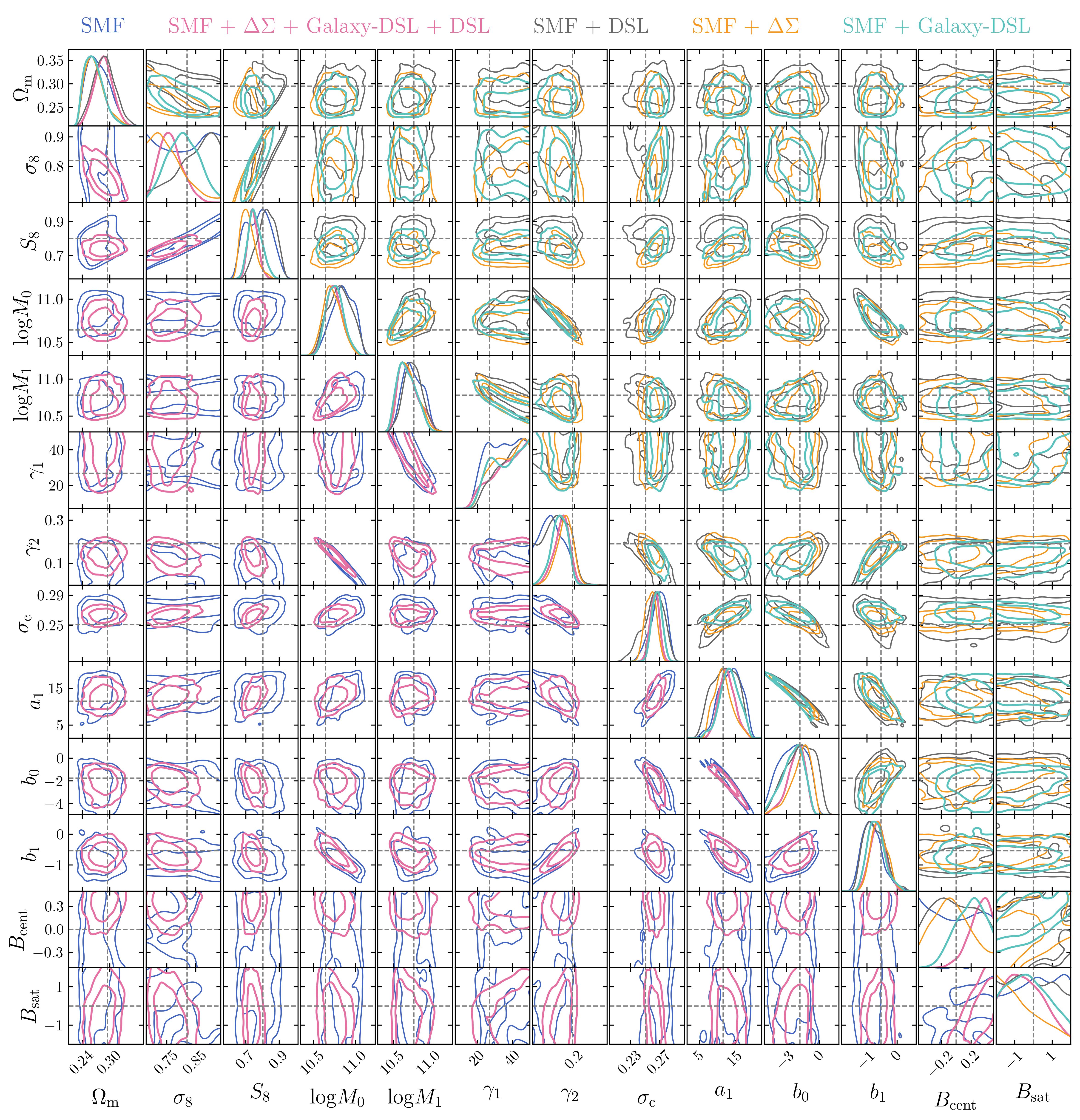}
\caption{Posterior distribution using the \citetalias{Takahashi2017} mocks as a reference and the emulator for the model for all the ESD statistics.}
\label{fig:MCMC_T17}
\end{figure*}

In order to still perform a T17 analysis, we use the mean over all 756 measurements as our reference data vector for the forecast analysis shown in Fig.~\ref{fig:MCMC_T17} for the different summary statistics. We would not necessarily expect an excellent fit due to the differences in mass resolutions and halo finder. Furthermore, \citetalias{Takahashi2017} assumes that the neutrinos are massless, but the \abacussummit\ have used $\omega_\mathrm{ncdm}=0.00064420$. This can have an $\sim 3\%$ effect on the matter power spectrum, especially on smaller scales \cite{Zennaro2019}, shifting the $\sigma_8$ up by $1.8\%$. However, since the \abacussummit\ is not necessarily a better fit to real data, using the \citetalias{Takahashi2017} simulations as a reference might indicate possible biases in real data analyses. In Fig.~\ref{fig:ESD_T17} and \ref{fig:SMF_T17}, we show the \texttt{pyccl} and emulator predictions compared to \citetalias{Takahashi2017} measurements that are used as the reference data vector. We can see that, especially for higher stellar mass bins, the match of the ESD at small scales is less accurate. 

If we consider only scales above $>8\,h^{-1}\mathrm{Mpc}$ the bias in the cosmological parameters is gone. The remaining inaccuracies at large scales are then compensated by adjusting especially the HOD parameters $B_\mathrm{cen}$ and $\sigma_\mathrm{c}$. Overall, we are not worried about modelling our summary statistics since these biases are at a 2-$\sigma$ level in high dimensional parameter space, and it is not clear which of the two simulations is the better match to real data. 

\bibliographystyle{apsrev4-1}
\bibliography{bibliography}
 
\end{document}